\documentclass[twocolumn,amsmath,amssymb]{revtex4}
\usepackage{amsmath,amssymb,graphicx}
\usepackage{graphicx}
\usepackage{dcolumn}
\usepackage{bm}
\usepackage{epsfig}
\usepackage{here}
\usepackage{float}
\usepackage{amsmath}
\usepackage[usenames]{color}
\usepackage{amsmath}
\usepackage{epsfig}
\usepackage{color}

\newcommand{\bea}{\begin{eqnarray}}
\newcommand{\eea}{\end{eqnarray}}

\begin{document}
%%%%%%%%%%%%%%%%%%%%%%%%%%%%%%%%%%%%%%%%%%%%%%%%%%%%%%%%%%%%%%%
%\title{Cosmological axion in post-Newtonian approximation}
\title{Axion cosmology with post-Newtonian corrections}
%\title{Post-Newtonian axion}
\author{Jai-chan Hwang${}^{1}$, Hyerim Noh${}^{2}$}
\address{${}^{1}$Center for Theoretical Physics of the Universe,
         Institute for Basic Science (IBS), Daejeon, 34051, Republic of Korea
         \\
         ${}^{2}$Theoretical Astrophysics Group, Korea Astronomy and Space Science Institute, Daejeon, Republic of Korea
         }

%%%%%%%%%%%%%%%%%%%%%%%%%%%%%%%%%%%%%%%%%%%%%%%%%%%%%%%%%%%%%%%
%\date{\today}

%%%%%%%%%%%%%%%%%%%%%%%%%%%%%%%%%%%%%%%%%%%%%%%%%%%%%%%%%%%%%%%
\begin{abstract}

We present first-order post-Newtonian (1PN) approximations of a general imperfect fluid and of an axion as a coherently oscillating massive scalar field, both in the cosmological context. For the axion, using the Klein transformation and Madelung transformation we derive the Schr\"odinger and Madelung hydrodynamic formulations, respectively, in exact covariant way and to 1PN order. Complete sets of equations for the 1PN formulations are derived without fixing the temporal gauge condition. We study the linear instability in cosmology and a static limit for both fluid and axion; these are presented independently of the gauge condition to 1PN order, thus are naturally gauge-invariant.

\end{abstract}
%%%%%%%%%%%%%%%%%%%%%%%%%%%%%%%%%%%%%%%%%%%%%%%%%%%%%%%%%%%%%%%

%%%%%%%%%%%%%%%%%%%%%%%%%%%%%%%%%%%%%%%%%%%%%%%%%%%%%%%%%%%%%%%
\maketitle

%\tableofcontents
%%%%%%%%%%%%%%%%%%%%%%%%%%%%%%%%%%%%%%%%%%%%%%%%%%%%%%%%%%%%%%%
%
%
%
%%%%%%%%%%%%%%%%%%%%%%%%%%%%%%%%%%%%%%%%%%%%%%%%%%%%%%%%%%%%%%%
\section{Introduction}
                                   \label{sec:Introduction}

Post-Newtonian (PN) hydrodynamics is a consistent approximation of Einstein's gravity with Newtonian hydrodynamics appearing as the zeroth-order PN (0PN) in the $c \rightarrow \infty$ limit in some dimensionless combinations of the metric and energy-momentum variables \cite{Chandrasekhar-1965, Poisson-Will-2014}. Cosmological extension is possible with different temporal gauge conditions readily available \cite{Hwang-Noh-Puetzfeld-2008}. The background Friedmann equations are subtracted in cosmology, and the remaining deviations from the background can be consistently expanded to the PN orders.

The scalar field with various forms of potential is popularly used in physical cosmology as tools for variety of essential cosmological roles. For example, the inflation, dark matter and dark energy are often modeled by using the scalar field. However, with general forms of potential the scalar field does not allow the PN approximation. This is understandable as the general scalar field does not necessarily have the Newtonian (0PN) limit. The situation changes as we consider an axion.

As an axion, we consider a coherently oscillating stage of a massive scalar field; we may include a self-interaction term {\it assuming} that due to small coupling it does not interfere the coherent oscillation of the field. The cosmological axion is known to behave as a zero-pressure fluid, thus non-relativistic in both background and perturbations \cite{axion-CDM, axion-pert}, in fact, even to fully-nonlinear and exact perturbations \cite{Noh-Hwang-Park-2017}. The axion, being an oscillating scalar field, actually has a characteristic stress (both isotropic and anisotropic) reflecting the wave nature and uncertainty principle \cite{Madelung-1927, Bohm-1952}. For extremely small axion mass with macroscopic Compton wavelength its effect becomes cosmologically important \cite{Hu-Barkana-Gruzinov-2000, FDM-review}.

Here we study the first-order PN (1PN) approximation of the cosmological axion. The PN expansion differs from the relativistic perturbation theory. In the latter, all deviations from the Friedmann background in the metric and energy-momentum tensor are regarded as perturbation, and a consistent expansion is made for the perturbations order by order assuming that the perturbations are small. In the former, the remaining deviations after subtracting the Friedmann background are expanded in PN expansion by identifying dimensionless PN variables with $c^{-1}$ involved. The lowest expansion gives Newtonian limit, and the next order involving $c^{-2}$, like $GM/Rc^2$, $\Phi/c^2$, $v^2/c^2$, $p/\varrho c^2$, etc., gives the 1PN expansion; $M$, $R$, $\Phi$, $v$, $p$ and $\varrho$ are characteristic mass, length, gravitational potential, velocity, pressure and density, respectively. The perturbation theory is fully relativistic but applicable for small deviations (i.e., weakly nonlinear), whereas, the PN expansion is weakly relativistic but fully nonlinear. Thus, the two approximations (if available) are complementary to each other.

The weak gravity limit is yet another complementary approximation where the gravity is assumed to be weak (near 0PN) while considering fully-relativistic and nonlinear energy-momentum (thus, $\infty$PN) \cite{Hwang-Noh-2016}. This approximation is relevant in many astrophysical situations including cosmology; in observable universe, except for nearby compact objects like neutron stars and black holes, the gravity represented by a dimensionless metric parameter $\Phi/c^2$ is extremely small with typical value less than $10^{-5}$ \cite{Kim-2022-CP}. The weak gravity limit of a general scalar field (including axion) in cosmology is presented in \cite{Hwang-Noh-2022b} in the context of electrodynamics and magnetohydrodynamics.

The PN approximation is available for axion only after a transformation which was introduced by Klein in his way to derive the Schr\"odinger equation as the non-relativistic limit of the Klein-Gordon equation \cite{Klein-1926, Dirac-1979, Chavanis-Matos-2017}. By further applying a transformation by Madelung the Schr\"odinger equation leads to Madelung's hydrodynamic formulation of the system \cite{Madelung-1927}. Although the Klein transformation is suitable to derive the non-relativistic limit and the Madelung transformation originally applied to the Schr\"odinger equation, here we apply these to the relativistic Klein-Gordon equation in our way to derive the PN corrections.

As our 1PN fluid formulation is valid for a general imperfect fluid, we can also derive the axion hydrodynamic equations from the fluid equations using the PN order fluid quantities for the axion. In this work, the 1PN fluid and axion formulations are applied to the gravitational instability and to a static limit. Both are available {\it without} imposing the temporal gauge condition. Thus, we derive the Jeans scale and the equilibrium scale for an axion fluid in naturally gauge-invariant manner.

In Sec.\ \ref{sec:1PN-equations} we summarize the complete set of 1PN equations for the three (hydrodynamic, Schr\"odinger, and Madlung) formulations without imposing temporal gauge condition, thus in gauge-ready forms. The equations are derived in Appendices. In Secs.\ \ref{sec:instability} and \ref{sec:static} we apply the formulations to the gravitational instability and the static limit. Section \ref{sec:Discussion} is a discussion. Appendix \ref{sec:cov-formulation} presents covariant forms of fluid quantities under the Klein and Madelung transformations. The complete PN corrections for these three (fluid, Schr\"odinger and Madelung) formulations are derived in Appendix \ref{sec:1PN-approximation}.

%%%%%%%%%%%%%%%%%%%%%%%%%%%%%%%%%%%%%%%%%%%%%%%%%%%%%%%%%%%%%%%%%
%
%
%
%%%%%%%%%%%%%%%%%%%%%%%%%%%%%%%%%%%%%%%%%%%%%%%%%%%%%%%%%%%%%%%%%
\section{Cosmological 1PN equations}
                                          \label{sec:1PN-equations}

Here we summarize the 1PN equations of the three formulations. Derivations are presented in the Appendices.

The 1PN metric convention is \cite{Chandrasekhar-1965}
\bea
   & & ds^2 = - \left[ 1 - {1 \over c^2} 2 U
       + {1 \over c^4} \left( 2 U^2 - 4 \Upsilon \right) \right]
       c^2 d t^2
   \nonumber \\
   & & \qquad
       - {1 \over c^3} 2 a P_i c dt d x^i
       + a^2 \left( 1 + {1 \over c^2} 2 V \right)
       \delta_{ij} d x^i d x^j,
\eea
where $a(t)$ is the cosmic scalar factor, the index of $P_i$ is raised and lowered using $\delta_{ij}$ and its inverse, and to 1PN order we have $V = U$. The energy-momentum (thus fluid quantities) convention can be found in Eqs.\ (\ref{four-vector-1PN})-(\ref{Tab-fluid-1PN}). In the above metric convention we ignored the transverse-tracefree tensor-type metric, and imposed the spatial gauge conditions (without losing any generality and convenience) to make the spatial part of the metric simple, but we have not imposed the temporal gauge condition yet \cite{Hwang-Noh-Puetzfeld-2008}. Together with the temporal gauge condition to be introduced below, all remaining 1PN variables are spatially and temporally gauge-invariant \cite{Hwang-Noh-Puetzfeld-2008}.

The 1PN order equations will be presented without imposing the temporal gauge condition. The general temporal gauge condition can be written as \cite{Hwang-Noh-Puetzfeld-2008}
\bea
   & & {1 \over a} P^i_{\;\;,i} + n \dot U
       + m {\dot a \over a} U = 0,
   \label{PN-gauges}
\eea
with arbitrary real numbers $n$ and $m$. As the gauge condition we can choose any number for $n$ and $m$; $n = 3$ (Chandrasekhar, standard PN gauge, or maximal slicing), $n = 3 = m$ (uniform-expansion gauge), $n = 4$ (harmonic gauge), $n = 0 = m$ (transverse-shear gauge), etc. As these gauge conditions, together with the spatial gauge conditions we already have imposed in the metric, completely remove the gauge degrees of freedom, all remaining 1PN variables after imposing the gauge condition can be equivalently regarded as gauge-invariant, see \cite{Hwang-Noh-Puetzfeld-2008}. Later in Secs.\ \ref{sec:instability} and \ref{sec:static}, we will show that for the gravitational instability and in the static equilibrium limit, analyses are possible without imposing the gauge condition. This implies that these two analyses are naturally gauge-invariant.

\begin{widetext}
%%%%%%%%%%%%%%%%%%%%%%%%%%%%%%%%%%%%%%%%%%%%%%%%%%%%%%%%%%%%%%%%%
\subsection{1PN hydrodynamic equations}

The hydrodynamic conservation-equations and Einstein equation to 1PN order give
\bea
   & & \dot \varrho
       + 3 H \varrho
       + {1 \over a} ( \varrho v^i )_{,i}
       + {1 \over c^2} \bigg\{
       ( \varrho \Pi + \varrho v^2 )^{\displaystyle{\cdot}}
       + 3 H (\varrho \Pi + p )
   \nonumber \\
   & & \qquad
       + {1 \over a} \left[ \varrho v^i
       ( \Pi + v^2 - 3 U )
       + \varrho P^i + p v^i + Q^i + \Pi^i_j v^j \right]_{,i}
       + \varrho \left( 3 \dot U
       + {2 \over a} U_{,i} v^i
       + 4 H v^2 \right)
       \bigg\} = 0,
   \label{PN-E-conserv} \\
   & & {1 \over a^4} ( a^4 \varrho v_i )^{\displaystyle{\cdot}}
       + {1 \over a} \left( \varrho v^j v_i + p \delta^j_i
       + \Pi^j_i \right)_{,j}
       - {1 \over a} \varrho U_{,i}
       + {1 \over c^2} \bigg\{
       {1 \over a^4} \left\{ a^4 \left[
       \varrho v_i ( \Pi + v^2 + U ) + p v_i
       + Q_i + \Pi_{ij} v^j \right] \right\}^{\displaystyle{\cdot}}
   \nonumber \\
   & & \qquad
       + {1 \over a} \left[ \varrho v^j v_i
       ( \Pi + v^2 - 2 U ) + \varrho P^j v_i
       + p v^j v_i + Q^j v_i + Q_i v^j
       - 2 U \Pi^j_i \right]_{,j}
   \nonumber \\
   & & \qquad
       + 2 \dot U \varrho v_i
       + {2 \over a} U_{,j}
       ( \varrho v^j v_i + \Pi^j_i )
       - {2 \over a} \varrho \Upsilon_{,i}
       - {1 \over a} U_{,i}
       \left[ \varrho ( \Pi + 2 v^2 ) + p \right]
       + \varrho v^j P_{j,i}
       \bigg\} = 0,
   \label{PN-Mom-conserv} \\
   & & {\Delta \over a^2} U
       + 4 \pi G ( \varrho - \varrho_b )
       + {1 \over c^2} \bigg\{
       2 {\Delta \over a^2} \Upsilon
       + 3 \left( \ddot U + 3 H \dot U
       + 2 {\ddot a \over a} U \right)
       - {2 \over a^2} U \Delta U
   \nonumber \\
   & & \qquad
       + {1 \over a^2} ( a P^i_{\;\;,i} )^{\displaystyle{\cdot}}
       + 8 \pi G \left[ {1 \over 2} ( \varrho \Pi
       - \varrho_b \Pi_b )
       + \varrho v^2
       + {3 \over 2} ( p - p_b ) \right] \bigg\} = 0,
   \label{PN-1-U-F} \\
   & & ( \dot U + H U )_{,i}
       + {1 \over 4 a} ( P^k_{\;\;,ki} - \Delta P_i )
       = 4 \pi G \varrho a v_i,
   \label{PN-2-U-F}
\eea
where $H \equiv \dot a/a$. To the background order, we have Eqs.\ (\ref{BG-eqs}) and (\ref{BG-conservation}).

%%%%%%%%%%%%%%%%%%%%%%%%%%%%%%%%%%%%%%%%%%%%%%%%%%%%%%%%%%%%%%%%%
\subsection{1PN Schr\"odinger equations}

The Klein transformation is \cite{Klein-1926}
\bea
   & & \phi \equiv {\hbar \over \sqrt{2 m}}
       \left( \psi e^{-i \omega_c t}
       + \psi^* e^{i \omega_c t} \right),
   \label{K-transformation}
\eea
where $\phi$ is a real scalar field and $\psi$ is a complex wavefunction; $\omega_c \equiv m c^2/\hbar$ is the Compton frequency. The Schr\"odinger-Einstein equations to 1PN order give
\bea
   & & \left( {\Delta \over a^2}
       - 8 \pi \ell_s | \psi |^2 \right) \psi
       + {2 i m \over \hbar} \left( \dot \psi
       + {3 \over 2} H \psi \right)
       + {2 m^2 \over \hbar^2} U \psi
       + {1 \over c^2} \bigg[
       - \ddot \psi - 3 H \dot \psi
       - 2 U {\Delta \over a^2} \psi
   \nonumber \\
   & & \qquad
       + {2 i m \over \hbar} \left( 2 U \dot \psi
       + {1 \over a} P^i \psi_{,i} \right)
       + {i m \over \hbar}
       \left( {1 \over a} P^i_{\;\;,i}
       + 4 \dot U + 6 H U \right) \psi
       + {2 m^2 \over \hbar^2}
       \left( 2 \Upsilon + U^2 \right) \psi
       \bigg]
       = 0,
   \label{PN-Schrodinger} \\
   & & {\Delta \over a^2} U
       + 4 \pi G m ( |\psi|^2 - |\psi_b|^2 )
       + {1 \over c^2} \bigg\{
       2 {\Delta \over a^2} \Upsilon
       + 3 \left( \ddot U + 3 H \dot U
       + 2 {\ddot a \over a} U \right)
       - {2 \over a^2} U \Delta U
       + {1 \over a^2} ( a P^i_{\;\;,i} )^{\displaystyle{\cdot}}
   \nonumber \\
   & & \qquad
       + 8 \pi G {\hbar^2 \over m}
       \left[ - {1 \over 2 a^2} \left( \psi \Delta \psi^*
       + \psi^* \Delta \psi \right)
       + 6 \pi \ell_s ( |\psi|^4 - |\psi_b|^4 )
       \right] \bigg\} = 0,
   \label{PN-1-Schrodinger} \\
   & & ( \dot U + H U )_{,i}
       + {1 \over 4 a} ( P^k_{\;\;,ki} - \Delta P_i )
       = 2 \pi G i \hbar \left( \psi \psi^*_{,i}
       - \psi^* \psi_{,i} \right).
   \label{PN-2-Schrodinger}
\eea
To the background order, we have Eq.\ (\ref{BG-eqs}) with
\bea
   & & \varrho_b = m |\psi_b|^2, \quad
       \varrho_b \Pi_b = 3 p_b
       = {6 \pi \ell_s \hbar^2 \over m} |\psi_b|^4, \quad
       \dot \psi_b \psi_b^* + \psi_b \dot \psi_b^*
       + 3 {\dot a \over a} |\psi_b|^2 \left( 1
       - {4 \pi \ell_s \hbar^2 \over m^2 c^2} |\psi_b|^2 \right)
       = 0.
\eea

%%%%%%%%%%%%%%%%%%%%%%%%%%%%%%%%%%%%%%%%%%%%%%%%%%%%%%%%%%%%%%%%%
\subsection{1PN Madelung equations}

The Madelung transformation is \cite{Madelung-1927}
\bea
   & & \psi \equiv \sqrt{\varrho \over m} e^{i m u/\hbar}.
   \label{M-transformation}
\eea
The Madelung-Einstein equations to 1PN order give
\bea
   & & \dot \varrho + 3 H \varrho
       + {1 \over a} \nabla \cdot \left( \varrho {\bf u} \right)
       + {1 \over c^2} \left[
       - \left( \varrho \dot u \right)^{\displaystyle{\cdot}}
       - 3 H \varrho \dot u
       - 4 U {1 \over a} \nabla \cdot \left( \varrho {\bf u} \right)
       + 4 \varrho \dot U
       + {1 \over a} \nabla \cdot \left( \varrho {\bf P} \right)
       \right]
       = 0,
   \label{PN-E-conserv-axion} \\
   & & \dot {\bf u} + H {\bf u}
       + {1 \over a} {\bf u} \cdot \nabla {\bf u}
       - {1 \over a} \nabla U
       - {\hbar^2 \over 2 m^2} {1 \over a} \nabla
       \left( {1 \over a^2} {\Delta \sqrt{\varrho} \over \sqrt{\varrho}}
       - {8 \pi \ell_s \over m} \varrho \right)
   \nonumber \\
   & & \qquad
       + {1 \over c^2} {1 \over a}
       \nabla \left[ - {1 \over 2} \dot u^2
       - 2 \Upsilon + U^2
       - 2 U {\bf u}^2
       + {\bf P} \cdot {\bf u}
       + {\hbar^2 \over 2 m^2} \left(
       {\ddot {\sqrt{\varrho}} \over \sqrt{\varrho}}
       + 3 H {\dot {\sqrt{\varrho}} \over \sqrt{\varrho}}
       + 4 U {1 \over a^2}
       {\Delta \sqrt{\varrho} \over \sqrt{\varrho}}
       - 2 U {8 \pi \ell_s \over m} \varrho
       \right) \right]
       = 0,
   \label{PN-Mom-conserv-axion} \\
   & & {\Delta \over a^2} U
       + 4 \pi G ( \varrho - \varrho_b )
       + {1 \over c^2} \bigg\{
       2 {\Delta \over a^2} \Upsilon
       + 3 \left( \ddot U + 3 H \dot U
       + 2 {\ddot a \over a} U \right)
       - {2 \over a^2} U \Delta U
       + {1 \over a^2} ( a \nabla \cdot {\bf P} )^{\displaystyle{\cdot}}
   \nonumber \\
   & & \qquad
       + 8 \pi G \varrho {\bf u}^2
       + 8 \pi G {\hbar^2 \over m^2}
       \left[ - {1 \over a^2} \sqrt{\varrho} \Delta \sqrt{\varrho}
       + {6 \pi \ell_s \over m} (\varrho^2 - \varrho_b^2)
       \right] \bigg\} = 0,
   \label{PN-1-axion} \\
   & & \nabla ( \dot U + H U )
       + {1 \over 4 a} ( \nabla \nabla \cdot {\bf P}
       - \Delta {\bf P} )
       = 4 \pi G \varrho a {\bf u},
   \label{PN-2-axion}
\eea
where we defined ${\bf u} \equiv {1 \over a} \nabla u$. The original equation of Eq.\ (\ref{PN-Mom-conserv-axion}) is
\bea
   & & \dot u - U + {1 \over 2} {\bf u}^2
       - {\hbar^2 \over 2 m^2} \left( {1 \over a^2}
       {\Delta \sqrt{\varrho} \over \sqrt{\varrho}}
       - {8 \pi \ell_s \over m} \varrho \right)
   \nonumber \\
   & & \qquad
       + {1 \over c^2} \left[ - {1 \over 2} \dot u^2
       - 2 \Upsilon + U^2
       - 2 U {\bf u}^2
       + {\bf P} \cdot {\bf u}
       + {\hbar^2 \over 2 m^2} \left(
       {\ddot {\sqrt{\varrho}} \over \sqrt{\varrho}}
       + 3 H {\dot {\sqrt{\varrho}} \over \sqrt{\varrho}}
       + 4 U {1 \over a^2}
       {\Delta \sqrt{\varrho} \over \sqrt{\varrho}}
       - 2 U {8 \pi \ell_s \over m} \varrho
       \right) \right]
       = 0.
   \label{dot-u}
\eea
The relation between $u$ and fluid velocity $v_i$, introduced in the four-vector in Eq.\ (\ref{four-vector-1PN}), is given in Eq.\ (\ref{u-v-relation}). To the background order, we have Eq.\ (\ref{BG-eqs}) with
\bea
   & & \varrho_b \Pi_b = 3 p_b
       = {6 \pi \ell_s \hbar^2 \over m^3} \varrho_b^2, \quad
       \dot \varrho_b + 3 {\dot a \over a} \varrho_b
       \left( 1 - {4 \pi \ell_s \hbar^2 \over m^3 c^2} \varrho_b
       \right)
       = 0.
\eea

\end{widetext}
%%%%%%%%%%%%%%%%%%%%%%%%%%%%%%%%%%%%%%%%%%%%%%%%%%%%%%%%%%%%%%%%%
\subsection{0PN equations}

To 0PN order, we take $c \rightarrow \infty$ limit. For a hydrodynamic fluid, the mass and momentum conservation equations and Poisson's equation are
\bea
   & & \dot \varrho
       + 3 H \varrho
       + {1 \over a} \nabla \cdot ( \varrho {\bf v} ) = 0,
   \label{0PN-E-conserv} \\
   & & {1 \over a^4} ( a^4 \varrho {\bf v}
       )^{\displaystyle{\cdot}}
       + {1 \over a} \left( \varrho v^j v_i + p \delta^j_i
       + \Pi^j_i \right)_{,j}
       - {1 \over a} \varrho \nabla U = 0,
   \nonumber \\
   \label{0PN-Mom-conserv} \\
   & & {\Delta \over a^2} U
       = - 4 \pi G ( \varrho - \varrho_b ).
   \label{0PN-Poisson}
\eea
For the axion, the Schr\"odinger-Poisson's equations are
\bea
   & & i \hbar \left( \dot \psi
       + {3 \over 2} H \psi \right)
       = -{\hbar^2 \over 2 m} \left( {\Delta \over a^2}
       - 8 \pi \ell_s | \psi |^2 \right) \psi
       - m U \psi,
   \nonumber \\
   \label{0PN-Schrodinger} \\
   & & {\Delta \over a^2} U = - 4 \pi G m \left( |\psi|^2
       - |\psi_b|^2 \right).
\eea
For the axion, the Madelung conservation equations with Poisson's equation are
\bea
   & & \dot \varrho + 3 H \varrho
       + {1 \over a} \nabla \cdot \left( \varrho {\bf u} \right)
       = 0,
   \label{0PN-E-conserv-axion} \\
   & & \dot {\bf u} + H {\bf u}
       + {1 \over a} {\bf u} \cdot \nabla {\bf u}
       - {1 \over a} \nabla U
   \nonumber \\
   & & \qquad
       - {\hbar^2 \over 2 m^2} {1 \over a} \nabla
       \left( {1 \over a^2} {\Delta \sqrt{\varrho} \over \sqrt{\varrho}}
       - {8 \pi \ell_s \over m} \varrho \right)
       = 0,
   \label{0PN-Mom-conserv-axion} \\
   & & {\Delta \over a^2} U
       = - 4 \pi G ( \varrho - \varrho_b ).
\eea
These equations are valid to fully nonlinear order.

%%%%%%%%%%%%%%%%%%%%%%%%%%%%%%%%%%%%%%%%%%%%%%%%%%%%%%%%%%%%%%%%%
%
%
%
%%%%%%%%%%%%%%%%%%%%%%%%%%%%%%%%%%%%%%%%%%%%%%%%%%%%%%%%%%%%%%%%%
\section{Gravitational instability}
                                          \label{sec:instability}

We consider linear perturbations in the cosmological context. We set
\bea
   & & \varrho \rightarrow \varrho + \delta \varrho
       = \varrho ( 1 + \delta), \quad
       \Pi \rightarrow \Pi + \delta \Pi,
   \nonumber \\
   & &
       p \rightarrow p + \delta p, \quad
       \psi \rightarrow \psi + \delta \psi.
\eea
The other variables are already perturbed order; we ignore the self-interaction term.
%, thus $\dot u = 0$ to the background order.
Here we keep only to linear order in perturbation variables, and in case clarification is needed we indicate the background order variables with a subindex $b$, like $\varrho_b$, etc.

%%%%%%%%%%%%%%%%%%%%%%%%%%%%%%%%%%%%%%%%%%%%%%%%%%%%%%%%%%%%%%%%%
\subsection{Hydrodynamics}

To the linear order perturbation, Eqs.\ (\ref{PN-E-conserv})-(\ref{PN-1-U-F}) give
\bea
   & & \left[ \varrho \left( 1 + {\Pi \over c^2} \right)
       \right]^{\displaystyle{\cdot}}
       + 3 {\dot a \over a} \left[ \varrho \left( 1 + {\Pi \over c^2} \right)
       + {p \over c^2} \right]
   \nonumber \\
   & & \qquad
       + {1 \over a} \nabla \cdot \left\{
       \left[ \varrho \left( 1 + {\Pi \over c^2} \right)
       + {p \over c^2} \right] {\bf v}
       + {1 \over c^2} \varrho {\bf P} \right\}
       + {3 \over c^2} \varrho \dot U
   \nonumber \\
   & & \qquad
       = 0,
   \label{PN-E-conserv-linear} \\
   & & {1 \over a^4} \left\{ a^4 \left\{
       \left[ \varrho \left( 1 + {\Pi \over c^2} \right)
       + {p \over c^2} \right] {\bf v}
       + {1 \over c^2} {\bf Q} \right\}
       \right\}^{\displaystyle{\cdot}}
       + {1 \over a} \nabla p
   \nonumber \\
   & & \qquad
       + {1 \over a} \Pi^j_{i,j}
       - {1 \over a} \nabla \left\{
       \left[ \varrho \left( 1 + {\Pi \over c^2} \right)
       + {p \over c^2} \right] U
       + {2 \over c^2} \varrho \Upsilon \right\}
   \nonumber \\
   & & \qquad
       = 0,
   \label{PN-Mom-conserv-linear} \\
   & & {\Delta \over a^2} \left( U
       + {2 \over c^2} \Upsilon \right)
   \nonumber \\
   & & \qquad
       + 4 \pi G \bigg[ \varrho
       \left( 1 + {\Pi \over c^2} \right)
       + {3 p \over c^2}
       - \varrho_b
       \left( 1 + {\Pi_b \over c^2} \right)
       - {3 p_b \over c^2}
       \bigg]
   \nonumber \\
   & & \qquad
       + {1 \over c^2} \left[ 3 \ddot U
       + 9 {\dot a \over a} \dot U
       + 6 {\ddot a \over a} U
       + {1 \over a^2} \left( a \nabla \cdot {\bf P}
       \right)^{\displaystyle{\cdot}} \right]
       = 0.
   \label{PN-1-U-F-linear}
\eea
Subtracting the background equation, we can derive density perturbation equation to 1PN order
\bea
  & & \ddot \delta_\mu
      + 2 {\dot a \over a} \dot \delta_\mu
      - {4 \pi G \mu \over c^2} \delta_\mu
      - {c^2 \over a^2 \mu} \left( \Delta \delta p
      + \Pi^{ij}_{\;\;\;,ij} \right)
   \nonumber \\
   & & \qquad
      + {1 \over c^2} \bigg\{
      {3 \over a^2} \left[ a^2 H
      \left( {\delta p \over \varrho}
      - {p \over \varrho} \delta \right)
      \right]^{\displaystyle{\cdot}}
      - 3 {\dot a \over a} {p \over \varrho} \dot \delta
   \nonumber \\
   & & \qquad
      - 4 \pi G \varrho \left( 3 {\delta p \over \varrho}
      - {p \over \varrho} \delta \right)
   \nonumber \\
   & & \qquad
      + {12 \pi G \varrho a^2 \over \Delta}
      \left[ {\dot a \over a} \dot \delta
      + \left( 2 {\ddot a \over a}
      - {\dot a^2 \over a^2} \right) \delta \right]
   \nonumber \\
   & & \qquad
      - {1 \over a^5 \varrho}
      \left( a^4 \nabla \cdot {\bf Q}
      \right)^{\displaystyle{\cdot}}
      \bigg\}
      = 0,
   \label{ddot-delta-eq-fluid}
\eea
where we used $\delta \equiv \delta \varrho/\varrho$ and $\delta_\mu \equiv \delta \mu/\mu = \delta + \delta \Pi/c^2$ which follows from $\mu \equiv \varrho (c^2 + \Pi)$; in this way, $\Pi_b$ and $\delta \Pi$ are absorbed to $\mu$ and $\delta_\mu$. In deriving this equation, as $P_i$ cancels we do not need to impose the gauge condition and Eq.\ (\ref{PN-2-U-F}) is not used. Thus, Eq.\ (\ref{ddot-delta-eq-fluid}) is naturally gauge-invariant. This is density perturbation equation valid to 1PN order; $p$, $\Pi_{ij}$ and $Q_i$ are provided by specifying the nature of the fluid; see below for the axion case.

The Jeans scale dividing the gravity and pressure dominating scales can be derived by setting $p = \delta \Pi = \Pi_{ij} = Q_i = 0$. In Fourier space with $\Delta = - k^2$ and introducing the sound velocity as $\delta p/\varrho \equiv v_s^2 \delta$, by setting the coefficients of $\delta$ terms equal to zero, assuming constant $v_s$, we can show
\bea
   {k_{\rm J} \over a}
       = {\sqrt{4 \pi G \varrho} \over v_s}
       \left( 1 - {1 \over \Omega} {v_s^2 \over c^2} \right),
   \label{k-Jeans-fluid}
\eea
where $\Omega \equiv 8 \pi G \varrho/(3 H^2)$. As we consider a general fluid, the axion can be regarded as a fluid. In order to derive the axion-Jeans scale to be derived later, we need to properly include the nonvanishing $\delta \Pi$ and $\dot v_s$, see below Eq.\ (\ref{k-Jeans-axion}).

For a zero-pressure ideal fluid ($Q_i = 0 = \Pi_{ij}$), we have
\bea
  & & \ddot \delta_\mu + 2 {\dot a \over a} \dot \delta_\mu
      - {4 \pi G \mu \over c^2} \delta_\mu
   \nonumber \\
   & & \qquad
      + {12 \pi G \varrho a^2 \over \Delta c^2}
      \left[ {\dot a \over a} \dot \delta
      + \left( 2 {\ddot a \over a}
      - {\dot a^2 \over a^2} \right) \delta \right]
      = 0.
   \label{ddot-delta-eq-fluid-ideal}
\eea
In the absence of internal energy we replace $\delta_\mu \rightarrow \delta$ and $\mu/c^2 \rightarrow \varrho$. This was derived in Eq.\ (25) of \cite{Noh-Hwang-2012}; comparison with relativistic linear perturbation was made in that paper. Equation (\ref{ddot-delta-eq-fluid-ideal}) coincides with the 1PN limit of relativistic density perturbation equation in zero-shear gauge and uniform-expansion gauge both of which show proper Newtonian limit in the sub-horizon limit for density and velocity perturbations and gravitational potential \cite{Hwang-Noh-1999}. The PN correction terms become important near horizon-scale while the PN expansion is reliable in sub-horizon scale. In the comoving gauge and the synchronous gauge Eq.\ (\ref{ddot-delta-eq-fluid-ideal}) {\it without} the PN correction is exactly valid in all scales \cite{Lifshitz-1946, Bardeen-1980}. However, in these two gauge conditions we cannot properly identify the gravitational potential.

%%%%%%%%%%%%%%%%%%%%%%%%%%%%%%%%%%%%%%%%%%%%%%%%%%%%%%%%%%%%%%%%%
\subsection{Axion hydrodynamics}

To the background order, Eqs.\ (\ref{PN-E-conserv-axion}) and (\ref{dot-u}) give
\bea
   & & \dot \varrho + 3 H \varrho = 0,
   \\
   & & \dot u = - {\hbar^2 \over 2 m^2 c^2} \bigg(
       {\ddot {\sqrt{\varrho}} \over \sqrt{\varrho}}
       + 3 H {\dot {\sqrt{\varrho}} \over \sqrt{\varrho}} \bigg)
   \nonumber \\
   & & \qquad
       = {3 \hbar^2 \over 4 m^2 c^2} \left( \dot H + {3 \over 2} H^2 \right)
       = {3 \hbar^2 \over 8 m^2} \Lambda.
\eea
Thus, $\varrho \propto a^{-3}$.

To the linear order perturbation, Eqs.\ (\ref{PN-E-conserv-axion})-(\ref{PN-1-axion}) using Eq.\ (\ref{dot-u}) give
\bea
   & & \dot \delta
       + {1 \over a} \nabla \cdot {\bf u}
   \nonumber \\
   & & \qquad
       + {1 \over c^2} \left[ 3 \dot U
       + {1 \over a} \nabla \cdot {\bf P}
       - {\hbar^2 \Delta \over 4 m^2 a^2}
       \left( \dot \delta - 2 {\dot a \over a} \delta
       \right) \right]
   \nonumber \\
   & & \qquad
       = 0,
   \label{E-conserv-linear} \\
   & & \dot {\bf u} + {\dot a \over a} {\bf u}
       - {1 \over a} \nabla U
       - {\hbar^2 \Delta \over 4 m^2 a^3} \nabla \delta
   \nonumber \\
   & & \qquad
       + {1 \over c^2} {1 \over a}
       \nabla \left( - 2 \Upsilon
       + {\hbar^2 \over 4 m^2} \ddot \delta \right)
       = 0,
   \label{Mom-conserv-linear} \\
   & & {\Delta \over a^2} U
       + 4 \pi G \varrho \delta
       + {1 \over c^2} \bigg[
       2 {\Delta \over a^2} \Upsilon
       + 3\ddot U + 9 {\dot a \over a} \dot U
       + 6 {\ddot a \over a} U
   \nonumber \\
   & & \qquad
       + {1 \over a^2} ( a \nabla \cdot {\bf P} )^{\displaystyle{\cdot}}
       - 4 \pi G \varrho {\hbar^2 \Delta \over m^2 a^2} \delta
       \bigg] = 0.
   \label{PN-1-U-linear}
\eea
Equation (\ref{PN-2-axion}) is not needed. Without imposing temporal gauge condition we can derive
\bea
   & & \ddot \delta + 2 {\dot a \over a} \dot \delta
       - 4 \pi G \varrho \delta
       + {\hbar^2 \Delta^2 \over 4 m^2 a^4} \delta
   \nonumber \\
   & & \qquad
       + {1 \over c^2} \bigg\{
       {12 \pi G \varrho a^2 \over \Delta}
       \left[ {\dot a \over a} \dot \delta
       + \left( 2 {\ddot a \over a}
       - {\dot a^2 \over a^2} \right) \delta \right]
   \nonumber \\
   & & \qquad
       + {\hbar^2 \Delta \over 2 m^2 a^2} \left(
       3 {\dot a \over a} \dot \delta
       + {\hbar^2 \Delta^2 \over 4 m^2 a^4} \delta \right)
       \bigg\}
       = 0.
   \label{ddot-delta-eq-axion}
\eea
These equations can also be derived from Eqs.\ (\ref{PN-E-conserv-linear})-(\ref{ddot-delta-eq-fluid}) by using the axion fluid quantities; to the linear order, ignoring the self-interaction terms, Eq.\ (\ref{fluid-axion-1PN}) gives
\bea
   & & \Pi_b = 0, \quad
       {p_b \over \varrho_b}
       = - {3 \hbar^2 \over 4 m^2 c^2} \dot H,
   \nonumber \\
   & &
       {\delta p \over \varrho}
       = - {\hbar^2 \Delta \over 4 m^2 a^2} \delta
       - {\hbar^2 \over 4 m^2 c^2} \left( \ddot \delta
       + 3 H \dot \delta + 3 \dot H \delta \right),
   \nonumber \\
   & &
       \delta \Pi = - {\hbar^2 \Delta \over 4 m^2 a^2} \delta, \quad
       Q_i = 0 = \Pi_{ij}.
   \label{fluid-axion-instability}
\eea
The background pressure $p_b$ appearing with $c^{-2}$ factor does not have any role to the 1PN order. Compared with the zero-pressure ideal fluid in Eq.\ (\ref{ddot-delta-eq-fluid-ideal}), the axion equation in Eq.\ (\ref{ddot-delta-eq-axion}) differs only in the quantum stress (with $\hbar^2$) terms appearing in both 0PN and 1PN orders.

From the coefficients of $\delta$-terms in Eq.\ (\ref{ddot-delta-eq-axion}), setting $\Delta \rightarrow - k^2$, we have the Jeans wavenumber
\bea
   {k_{\rm J} \over a}
       = 2 \sqrt{\sqrt{\pi G \varrho} m \over \hbar }
       \left[ 1 - \left( 1 - {1 \over 2 \Omega} \right)
       {\sqrt{\pi G \varrho} \hbar \over m c^2} \right].
   \label{k-Jeans-axion}
\eea
Thus, the 1PN correction is of $\sqrt{G \varrho}/\omega_c \sim H / \omega_c = \hbar H/(m c^2)$ order. To 1PN order this differs from the fluid case in Eq.\ (\ref{k-Jeans-fluid}). As mentioned below Eq.\ (\ref{k-Jeans-fluid}), we can derive this result directly from Eq.\ (\ref{ddot-delta-eq-fluid}) using Eq.\ (\ref{fluid-axion-instability}) and
\bea
   v_s^2 = {\hbar^2 \over 4 m^2} \left( {k^2 \over a^2}
       + {12 \pi G \varrho \over c^2} \right).
\eea

%%%%%%%%%%%%%%%%%%%%%%%%%%%%%%%%%%%%%%%%%%%%%%%%%%%%%%%%%%%%%%%%%
\subsection{Schr\"odinger formulation}

From the Madeluing transformation, we have
\bea
   & & \varrho = m |\psi|^2, \quad
       u = {\hbar \over 2 i m} \ln{(\psi/\psi^*)},
\eea
to the background, and
\bea
   & & \delta = {\delta \psi \over \psi}
       + {\delta \psi^* \over \psi^*}, \quad
       {2 i m \over \hbar} \delta u = {\delta \psi \over \psi}
       - {\delta \psi^* \over \psi^*},
\eea
to perturbations. To the background order, Eq.\ (\ref{PN-Schrodinger}) gives
\bea
   & & \dot \psi + {3 \over 2} {\dot a \over a} \psi
       - {\hbar \over 2 i m c^2}
       \left( \ddot \psi + 3 {\dot a \over a} \dot \psi \right)
       = 0.
\eea
%Using $\varrho \Pi = 0 = p$ to the background order, we have $\dot \varrho + 3 H \varrho = 0$ even to 1PN order.
Using
\bea
   & & {\delta \psi \over \psi}
       = {1 \over 2} \delta + {i m \over \hbar} \delta u,
\eea
the imaginary and real parts of Eq.\ (\ref{PN-Schrodinger}) give
\bea
   & & \dot \delta + {\Delta \over a^2} \delta u
       + {1 \over c^2} \left[ 3 \dot U
       + {1 \over a} P^i_{\;\;,i}
       - {\hbar^2 \Delta \over 4 m^2 a^2}
       \left( \dot \delta - 2 {\dot a \over a} \delta \right)
       \right]
   \nonumber \\
   & & \qquad
       = 0,
   \\
   & & \delta \dot u - U
       - {\hbar^2 \Delta \over 4 m^2 a^2} \delta
       + {1 \over c^2} \left(
       {\hbar^2 \over 4 m^2} \ddot \delta - 2 \Upsilon \right)
       = 0.
\eea
Using ${\bf u} \equiv {1 \over a} \nabla u$, we have Eqs.\ (\ref{E-conserv-linear}) and (\ref{Mom-conserv-linear}), and Eq.\ (\ref{PN-1-Schrodinger}) gives Eq.\ (\ref{PN-1-U-linear}).

%%%%%%%%%%%%%%%%%%%%%%%%%%%%%%%%%%%%%%%%%%%%%%%%%%%%%%%%%%%%%%%%%
%
%
%
%%%%%%%%%%%%%%%%%%%%%%%%%%%%%%%%%%%%%%%%%%%%%%%%%%%%%%%%%%%%%%%%%
\section{Static limit}
                                          \label{sec:static}

We consider the static limit with ${\bf v} = \dot \varrho = 0$, etc., in Minkowski background, thus $a \equiv 1$, $\Lambda = \varrho_b = 0$, etc. We ignore the self-interaction.

%%%%%%%%%%%%%%%%%%%%%%%%%%%%%%%%%%%%%%%%%%%%%%%%%%%%%%%%%%%%%%%%%
\subsection{Hydrodynamics}

For the fluid, without imposing the gauge condition, Eqs.\ (\ref{PN-Mom-conserv}) and (\ref{PN-1-U-F}) give
\bea
   & & \nabla \left( U + {1 \over c^2} 2 \Upsilon \right)
   \nonumber \\
   & & \qquad
       = {1 \over \varrho ( 1 + {1 \over c^2} \Pi )
       + {p \over c^2}}
       \left[ \nabla p
       + \left( 1 - {1 \over c^2} 2 U \right)
       \Pi^j_{i,j} \right],
   \label{Euler-static-fluid} \\
   & & \Delta \left( U + {1 \over c^2} 2 \Upsilon \right)
   \nonumber \\
   & & \qquad
       = - 4 \pi G \left[ \varrho \left( 1 + {1 \over c^2} \Pi
       + {1 \over c^2} 2 U \right)
       + 3 {p \over c^2} \right].
   \label{Poisson-static-fluid}
\eea
Combining the above equations, we have
\bea
   \nabla \cdot \left( {\nabla p
       + ( 1 - {1 \over c^2} 2 U ) \Pi^j_{i,j}
       \over \mu + p} \right)
       = - {4 \pi G \over c^4}
       {\mu + 3 p \over 1 - {1 \over c^2} 2 U},
   \label{EQ-static-fluid}
\eea
which is consistent with the Oppenheimer-Volkoff equation in spherically symmetric case to the 1PN order \cite{Tolman-1939}. As $P_i$ disappears in the above equations, all variables are naturally gauge-invariant, and we do not need Eqs.\ (\ref{PN-E-conserv}) and (\ref{PN-2-U-F}) which give $(\varrho P^i)_{,i} = 0$ and $P^k_{\;\;,ki} = \Delta P_i$ to 1PN order.

%%%%%%%%%%%%%%%%%%%%%%%%%%%%%%%%%%%%%%%%%%%%%%%%%%%%%%%%%%%%%%%%%
\subsection{Axion hydrodynamics}

For an axion, combining Eqs.\ (\ref{PN-Mom-conserv-axion}) and (\ref{PN-1-axion}), we have
\bea
   & & {\hbar^2 \Delta \over 2 m^2}
       \left[ \left( 1 + {3 \hbar^2 \over 2 m^2 c^2}
       {\Delta \sqrt{\varrho} \over \sqrt{\varrho}} \right)
       {\Delta \sqrt{\varrho} \over \sqrt{\varrho}} \right]
   \nonumber \\
   & & \qquad
       = 4 \pi G \varrho \left( 1
       - {3 \hbar^2 \over m^2 c^2}
       {\Delta \sqrt{\varrho} \over \sqrt{\varrho}} \right).
   \label{EQ-static-axion}
\eea
This is valid independently of the gauge condition, and also follows from Eq.\ (\ref{EQ-static-fluid}) using the fluid quantities in Eq.\ (\ref{fluid-axion-1PN}).

By setting $\Delta \sqrt{\varrho}/\sqrt{\varrho} \rightarrow - k^2/2$ ($1/2$-factor to match with the Jeans scale) we have the equilibrium wavenumber
\bea
   & & k_{\rm EQ}
       = 2 \sqrt{\sqrt{\pi G \varrho} m \over \hbar }
       \left( 1 + {9 \over 4}
       {\sqrt{\pi G \varrho} \hbar \over m c^2} \right).
\eea
This can be compared with Jeans wavenumber in Eq.\ (\ref{k-Jeans-axion}).

%%%%%%%%%%%%%%%%%%%%%%%%%%%%%%%%%%%%%%%%%%%%%%%%%%%%%%%%%%%%%%%%%
\subsection{Schr\"odinger formulation}

For an axion, combining the real part of Eq.\ (\ref{PN-Schrodinger}) and Eq.\ (\ref{PN-1-Schrodinger}), we have
\bea
   & & {\hbar^2 \Delta \over 2 m^2}
       \left[ \left( 1 + {3 \hbar^2 \over 2 m^2 c^2}
       {\Delta \psi \over \psi} \right)
       {\Delta \psi \over \psi} \right]
   \nonumber \\
   & & \qquad
       = 4 \pi G \varrho \left( 1
       - {3 \hbar^2 \over m^2 c^2}
       {\Delta \psi \over \psi} \right),
   \label{EQ-static-Schrodinger}
\eea
which is valid independently of the gauge condition. The imaginary part gives $(P^i \psi^2 )_{,i} = 0$. For $u_{,i} = 0$, we have $\Delta \psi/\psi = \Delta \sqrt{\varrho}/\sqrt{\varrho}$, and Eq.\ (\ref{EQ-static-Schrodinger}) leads to Eq.\ (\ref{EQ-static-axion}).

%%%%%%%%%%%%%%%%%%%%%%%%%%%%%%%%%%%%%%%%%%%%%%%%%%%%%%%%%%%%%%%%%
%
%
%
%%%%%%%%%%%%%%%%%%%%%%%%%%%%%%%%%%%%%%%%%%%%%%%%%%%%%%%%%%%%%%%%%
\section{Discussion}
                                          \label{sec:Discussion}

The PN approximation, being weakly relativistic but fully nonlinear, provides a complementary method to the relativistic perturbation theory which is fully relativistic but weakly nonlinear. Here we presented complete sets of 1PN approximation equations for a general imperfect fluid and an axion in the cosmological context. In the axion case we present the Schr\"odinger and Madelung hydrodynamic formulations where PN expansions are possible. All PN formulations are derived without fixing the temporal gauge condition. The complete sets of equations for the three formulations are summarized in Sec.\ \ref{sec:1PN-equations}. Detailed derivations are presented in two Appendices; these include the covariant formulations and the 1PN approximations for the axion. As applications we studied the gravitational instability and a static limit of the 1PN formulations. We analyzed these two cases without imposing the gauge condition, thus results are naturally gauge-invariant, see Secs.\ \ref{sec:instability} and \ref{sec:static}.

In \cite{Noh-Hwang-Park-2017} we derived relativistic axion density perturbation equation valid to fully-nonlinear order by using the fully-nonlinear and exact perturbation formulation made for a fluid \cite{Hwang-Noh-2013}. We took the axion-comoving gauge setting time-average of the longitudinal part of $T^0_i$ equals to zero; we note that although we have not imposed the temporal gauge condition in our study of the PN order gravitational instability in Sec.\ \ref{sec:instability}, the PN approximation does not allow the comoving gauge condition which implies vanishing perturbed lapse function, $\alpha$ (the Newtonian gravitational potential) in Eq.\ (\ref{metric-pert}), for a zero-pressure medium. In that study, by {\it assuming} $H/\omega_c \ll 1$ we arrived at the same equation known in non-relativistic limit except for relativistic contributions from the metric. By strictly ignoring $H/\omega_c$ higher order term, which is $\hbar H/(mc^2)$ thus 1PN order, \cite{Noh-Hwang-Park-2017} has derived the non-relativistic limit of the axion part. In this work we presented the 1PN extension. The non-relativistic (0PN) and 1PN approximations of hydrodynamics as limits of the relativistic fully-nonlinear perturbation formulation were presented in \cite{Hwang-Noh-2013-Newtonian-PN}.

The PN approximation can be applied to situations where all relativistic effects are small but not negligible. The 1PN equations are fully nonlinear, and the equations are designed so that the relativistic effects appear as the PN correction terms in the more familiar Newtonian hydrodynamic equations. Thus, the PN formulation is easier for numerical simulations compared with the full-blown numerical relativity. By setting $a \equiv 1$, ignoring the background fluid quantities and $\Lambda$, the formulations are valid in the Minkowski background. The 1PN equations are generally valid and can be applied to any astrophysical system where a single component fluid or axion is dominating. Extension to multi-component fluid in combination with axion is trivial, see \cite{Hwang-Noh-Park-2016} for multi-component fluids and scalar fields.

%%%%%%%%%%%%%%%%%%%%%%%%%%%%%%%%%%%%%%%%%%%%%%%%%%%%%%%%%%%%%%%
%
%
%
%%%%%%%%%%%%%%%%%%%%%%%%%%%%%%%%%%%%%%%%%%%%%%%%%%%%%%%%%%%%%%%
%\section*{Acknowledgments}
\vskip .5cm \centerline{\bf Acknowledgments}

H.N.\ was supported by the National Research Foundation (NRF) of Korea funded by the Korean Government (No.\ 2018R1A2B6002466 and No.\ 2021R1F1A1045515). J.H.\ was supported by IBS under the project code, IBS-R018-D1, and by the NRF of Korea funded by the Korean Government (No.\ NRF-2019R1A2C1003031).

%%%%%%%%%%%%%%%%%%%%%%%%%%%%%%%%%%%%%%%%%%%%%%%%%%%%%%%%%%%%%%%%%
%
%
%
%%%%%%%%%%%%%%%%%%%%%%%%%%%%%%%%%%%%%%%%%%%%%%%%%%%%%%%%%%%%%%%%%

%%%%%%%%%%%%%%%%%%%%%%%%%%%%%%%%%%%%%%%%%%%%%%%%%%%%%%%%%%%%%%%

\appendix
%%%%%%%%%%%%%%%%%%%%%%%%%%%%%%%%%%%%%%%%%%%%%%%%%%%%%%%%%%%%%%%%%
%
%
%
%%%%%%%%%%%%%%%%%%%%%%%%%%%%%%%%%%%%%%%%%%%%%%%%%%%%%%%%%%%%%%%%%
\section{Covariant formulation}
                                 \label{sec:cov-formulation}

We present covariant equations of the axion under the Klein and Madelung transformations. Although both transformations are applicable in the non-relativistic limit, here we apply these in the relativistic and covariant level. The covariant ($1 + 3$) equations for a general fluid can be found in \cite{Ellis-1971}. In the case of axion, what we need are fluid quantities of the axion and the equation of motion replacing (or complementing) the energy and momentum conservation equations. We will present these for the Schr\"odinger and Madelung formulations of the axion.

%%%%%%%%%%%%%%%%%%%%%%%%%%%%%%%%%%%%%%%%%%%%%%%%%%%%%%%%%%%%%%%%%
\subsection{Scalar field}

The fluid quantities are introduced based on a time-like four-vector $u_a$, normalized with $u^a u_a \equiv -1$, as \cite{Ellis-1971}
\bea
   T_{ab} = \mu u_a u_b
       + p \left( g_{ab} + u_a u_b \right)
       + q_a u_b + q_b u_a + \pi_{ab},
   \label{Tab}
\eea
where $\mu$, $p$, $q_a$ and $\pi_{ab}$ are the energy density, pressure, energy flux and anisotropic stress, respectively, with $q_a u^a \equiv 0 \equiv \pi_{ab} u^b$, $\pi_{ab} = \pi_{ba}$, and $\pi^a_a \equiv 0$.
Thus, we have
\bea
   & & \mu = T_{ab} u^a u^b, \quad
       p = {1 \over 3} T_{ab} h^{ab}, \quad
       q_a = - T_{cd} u^c h^d_a,
   \nonumber \\
   & &
       \pi_{ab} = T_{cd} h^c_a h^d_b - p h_{ab},
   \label{fluid-quantities}
\eea
where $h_{ab} \equiv g_{ab} + u_a u_b$ is the spatial-projection tensor. The fluid quantities have $13$ independent components ($\mu$, $p$, three $u_a$, three $q_a$ and five $\pi_{ab}$), whereas $T_{ab}$ needs only 10 independent components. Thus, we can freely impose three frame-conditions without any physical constraint. Often used ones are the normal-frame setting $u_i \equiv 0$, thus $u_a = n_a$, and the energy-frame setting $q_i \equiv 0$, thus $q_a = 0$.

The energy and the momentum conservation equations follow from $u_a T^{ab}_{\;\;\;\;;b} = 0$ and $h^c_a T^{ab}_{\;\;\;\; ;b} = 0$, respectively
\bea
   & & \widetilde {\dot {\mu}}
       + \left( \mu + p \right) \theta
       + \pi^{ab} \sigma_{ab}
       + q^a_{\;\;;a} + q^a a_a = 0,
   \label{cov-E-conserv-App} \\
   & & \left( \mu + p \right) a_a
       + h^b_a \left( p_{,b} + \pi^c_{b;c}
       + \widetilde {\dot {q}}_b \right)
   \nonumber \\
   & & \qquad
       + \left( \omega_{ab} + \sigma_{ab}
       + {4 \over 3} \theta h_{ab} \right) q^b = 0,
   \label{cov-Mom-conserv-App}
\eea
where $\widetilde {\dot \mu} \equiv \mu_{,c} u^c$; the expansion scalar ($\theta$), the acceleration vector ($a_a$), the rotation tensor ($\omega_{ab}$), and the shear tensor ($\sigma_{ab}$) are introduced as
\bea
   & &
       \theta \equiv u^a_{\;\; ;a}, \quad
       a_a \equiv {\widetilde {\dot {u}}}_a
       \equiv u_{a;b} u^b, \quad
       \sigma_{ab} \equiv \theta_{ab}
       - {1 \over 3} \theta h_{ab},
   \nonumber \\
   & & h^c_a h^d_b u_{c;d}
       = h^c_{[a} h^d_{b]} u_{c;d}
       + h^c_{(a} h^d_{b)} u_{c;d}
       \equiv \omega_{ab} + \theta_{ab},
   \label{kinematic-shear}
\eea
with $A_{[ab]} \equiv {1 \over 2} (A_{ab} - A_{ba})$ and $A_{(ab)} \equiv {1 \over 2} (A_{ab} + A_{ba})$.

We consider a minimally coupled scalar field in Einstein's gravity. We choose our convention in the Lagrangian density as
\bea
   {\cal L} = \sqrt{-g} \left[ {c^4 \over 16 \pi G}
       \left( R - 2 \Lambda \right)
       - {1 \over 2} \phi^{;c} \phi_{,c} - V (\phi)
       + L_{\rm m} \right],
\eea
where $L_{\rm m}$ is the matter part Lagrangian and $\Lambda$ is the cosmological constant. For the scalar field, the equation of motion and the energy-momentum tensor are
\bea
   & & \Box \phi = V_{,\phi},
   \label{EOM-MSF} \\
   & & T_{ab}
       = \phi_{,a} \phi_{,b}
       - \left( {1 \over 2} \phi^{;c} \phi_{,c}
       + V \right) g_{ab}.
   \label{Tab-MSF}
\eea
The fluid quantities in Eq.\ (\ref{fluid-quantities}) give
\bea
   & & \mu = {1 \over 2} \widetilde {\dot \phi}{}^2
       + V + {1 \over 2} h^{ab} \phi_{,a} \phi_{,b},
   \nonumber \\
   & &
       p = {1 \over 2} \widetilde {\dot \phi}{}^2
       - V - {1 \over 6} h^{ab} \phi_{,a} \phi_{,b}, \quad
       q_a = - \widetilde {\dot \phi} h^b_a \phi_{,b},
   \nonumber \\
   & &
       \pi_{ab} = h_a^c \phi_{,c} h_b^d \phi_{,d}
       - {1 \over 3} h_{ab} h^{cd} \phi_{,c} \phi_{,d}.
   \label{fluid-MSF-cov}
\eea
The equation of motion in Eq.\ (\ref{EOM-MSF}) gives
\bea
   \widetilde {\ddot \phi}
       + \theta \widetilde {\dot \phi}
       + V_{,\phi}
       - h_a^b \left( h^{ac} \phi_{,c} \right)_{;b}
       - h_a^b \phi_{,b} a^a
       = 0.
   \label{EOM-MSF-cov}
\eea

By taking the energy-frame condition, $q_a \equiv 0$, we have $h_a^b \phi_{,b} = 0$, thus $u_a = - \phi_{,a}/\widetilde {\dot \phi}$, and the fluid quantities and the equation of motion are simplified as
\bea
   & & \mu = {1 \over 2} \widetilde {\dot \phi}{}^2 + V, \quad
       p = {1 \over 2} \widetilde {\dot \phi}{}^2 - V, \quad
       \pi_{ab} = 0,
   \label{fluid-MSF-E-frame} \\
   & & \widetilde {\ddot \phi}
       + \theta \widetilde {\dot \phi}
       + V_{,\phi}
       = 0.
   \label{EOM-MSF-E-frame}
\eea
Using the fluid quantities in Eq.\ (\ref{fluid-MSF-E-frame}), we can show that Eq.\ (\ref{cov-E-conserv-App}) gives Eq.\ (\ref{EOM-MSF-E-frame}) and Eq.\ (\ref{cov-Mom-conserv-App}) is naturally valid.

The energy-frame condition $h_a^b \phi_{,b} = 0$ imposed in the field, however, is {\it not} suitable for the axion in coherent oscillation stage. In the axion case $h_i^b \phi_{,b} = 0$ is not necessarily the same as $\widetilde {\dot \phi} h^b_i \phi_{,b} = 0$; for example, $\pi_{ab} \neq 0$ in the axion case. The difference did not appear in the linear order perturbation studied in \cite{Hwang-Noh-2022-oscillation}. In general, we can use Eqs.\ (\ref{EOM-MSF}) and (\ref{fluid-MSF-cov}), instead.

%%%%%%%%%%%%%%%%%%%%%%%%%%%%%%%%%%%%%%%%%%%%%%%%%%%%%%%%%%%%%%%%%
\subsection{Relativistic Schr\"odinger formulation}
                                 \label{sec:K-transformation-cov}

The Klein transformation is in Eq.\ (\ref{K-transformation}). We consider a scalar field potential \cite{Brax-Valageas-Cembranos-2019}
\bea
   V = \sum_{n=1} {\lambda_{2n} \over 2n} \phi^{2n}
         = \sum_{n=1} {\lambda_{2n} \over 2n}
         {(2n)! \over (n!)^2}
         \left( {\hbar^2 |\psi|^2 \over 2 m} \right)^n,
\eea
where in the second step we used the Klein transformation and {\it ignored} (by time-averaging) oscillating terms. In the following we consider up to $n=2$, thus
\bea
   V = {1 \over 2} {m^2 c^2 \over \hbar^2} \phi^2
       + {1 \over 3} {m^2 \over \hbar^4} g \phi^4
       = {1 \over 2} m c^2 |\psi|^2
       + {1 \over 2} g |\psi|^4,
\eea
where we set $\lambda_2 \equiv {m^2 c^2 \over \hbar^2}$, $\lambda_4 \equiv {4 m^2 \over 3 \hbar^4} g$ and $g \equiv {4 \pi \ell_s \hbar^2 \over m}$.

\begin{widetext}
The Klein-Gordon equation and the energy-momentum tensor become \cite{Hwang-Noh-2022a}
\bea
   & & 0 = \Box \phi - V_{,\phi}
       = {\hbar \over \sqrt{2 m}} \bigg\{ e^{-i m c^2 t/\hbar} \left[ \Box \psi
       - {2 i m c \over \hbar} g^{0c} \psi_{,c}
       + {i m c \over \hbar} g^{ab} \Gamma^0_{ab} \psi
       - {m^2 c^2 \over \hbar^2}
       \left( g^{00} + 1 \right) \psi
       - {2 m \over \hbar^2} g |\psi|^2 \psi
       \right] + {\rm c.c.} \bigg\},
   \label{Schrodinger-eq} \\
   & & T_{ab} = m c^2 | \psi |^2 \delta^0_a \delta^0_b
       + i c \hbar
       \left( \psi_{,(a} \delta^0_{b)} \psi^*
       - \psi^*_{,(a} \delta^0_{b)} \psi \right)
       + {\hbar^2 \over m} \psi_{,(a} \psi^*_{,b)}
   \nonumber \\
   & & \qquad
       - {1 \over 2} g_{ab} \left[
       ( 1 + g^{00} ) m c^2 | \psi |^2
       + i c \hbar
       \left( \psi^{;0} \psi^* - \psi^{*;0} \psi \right)
       + {\hbar^2 \over m} \psi^{;c} \psi^*_{,c}
       + g |\psi|^4 \right],
   \label{Tab-MSF-psi}
\eea
where we {\it ignored} oscillating terms in $T_{ab}$, and used
\bea
   & & V_{,\phi}
       = {\partial \psi \over \partial \phi} V_{,\psi}
       + {\partial \psi^* \over \partial \phi} V_{,\psi^*}.
\eea
These can be derived in action formulation. The field part of Lagrangian is
\bea
   & & {\cal L}
       = - \sqrt{-g} \left[ {1 \over 2} \phi^{;c} \phi_{,c}
       + V(\phi) \right]
       = - \sqrt{-g} \left[
       {1 \over 2} m c^2 g^{00} |\psi|^2
       + {1 \over 2} i c \hbar
       ( \psi^{;0} \psi^* - \psi^{*;0} \psi )
       + {\hbar^2 \over 2m} \psi^{;c} \psi^*_{,c}
       + V(\psi, \psi^*) \right].
\eea
Variations with respect to $\phi$, $\psi$, $\psi^*$ and $g_{ab}$, with $\delta {\cal L} = {1 \over 2} \sqrt{-g} T^{ab} \delta g_{ab}$, lead to Eqs.\ (\ref{EOM-MSF}), (\ref{Tab-MSF}), (\ref{Schrodinger-eq}) and (\ref{Tab-MSF-psi}).

Thus, the equation of motion becomes
\bea
   & & \Box \psi
       - c g^{0c} {2 i m \over \hbar} \psi_{,c}
       + c g^{ab} \Gamma^0_{ab} {i m \over \hbar} \psi
       - c^2 \left( g^{00} + 1 \right) {m^2 \over \hbar^2} \psi
       - 8 \pi \ell_s | \psi |^2 \psi = 0.
   \label{relativistic-Schrodinger-eq}
\eea
This is a Schr\"odinger equation in the relativistic form. More properly, it is the Klein-Gordon equation written in terms of $\psi$, and in the absence of the self-interaction term, it leads to the Schr\"odinger equation in the non-relativistic limit with $c \rightarrow \infty$, see below.

Using Eqs.\ (\ref{fluid-quantities}) and (\ref{Tab-MSF-psi}), the fluid quantities become
\bea
   & & \mu = m |\psi|^2 c^2 \left[ u^0 u^0 + {1 \over 2}
       \left( g^{00} + 1 \right) \right]
       + {i \hbar} c \left( u^0 u^c
       + {1 \over 2} g^{0c} \right)
       \left( \psi_{,c} \psi^* - \psi^*_{,c} \psi \right)
       + {\hbar^2 \over m} \left(
       |\psi_{,c} u^c |^2
       + {1 \over 2} \psi^{;c} \psi^*_{,c}
       + 2 \pi \ell_s | \psi |^4 \right),
   \nonumber \\
   & & p = {1 \over 3} m |\psi|^2 c^2
       \left[ u^0 u^0 - {1 \over 2}
       \left( g^{00} + 3 \right) \right]
       + i \hbar {1 \over 3} c \left( u^0 u^c
       - {1 \over 2} g^{0c} \right)
       \left( \psi_{,c} \psi^* - \psi^*_{,c} \psi \right)
       + {1 \over 3} {\hbar^2 \over m} \left(
       |\psi_{,c} u^c |^2
       - {1 \over 2} \psi^{;c} \psi^*_{,c}
       - 6 \pi \ell_s | \psi |^4 \right),
   \nonumber \\
   & & q_a = - m |\psi|^2 c^2
       u^0 \left( u^0 u_a + c t_{,a} \right)
       - i \hbar c \left[
       {1 \over 2} \left( \psi_{,a} \psi^*
       - \psi^*_{,a} \psi \right) u^0
       + \left( \psi_{,c} \psi^* - \psi^*_{,c} \psi \right) u^c
       \left( u^0 u_a + {1 \over 2} c t_{,a} \right)
       \right]
   \nonumber \\
   & & \qquad
       - {\hbar^2 \over m} \left[
       |\psi_{,c} u^c |^2 u_a
       + {1 \over 2} \left( \psi_{,a} \psi^*_{,c}
       + \psi^*_{,a} \psi_{,c} \right) u^c \right],
   \nonumber \\
   & & \pi_{ab} = m |\psi|^2 c^2 \left[
       \left( u^0 u_a + c t_{,a} \right)
       \left( u^0 u_b + c t_{,b} \right)
       - {1 \over 3} \left( g_{ab} + u_a u_b \right)
       \left( g^{00} + u^0 u^0 \right) \right]
   \nonumber \\
   & & \qquad
       + i \hbar \bigg[
       c \left( u^0 u_{(a} + c t_{,(a} \right)
       \left( \psi_{,b)} \psi^* - \psi^*_{,b)} \psi \right)
       + c \left( u^0 u_{(a} + c t_{,(a} \right) u_{b)}
       \left( \psi_{,c} \psi^* - \psi^*_{,c} \psi \right) u^c
   \nonumber \\
   & & \qquad
       - {1 \over 3} \left( g_{ab} + u_a u_b \right)
       c \left( g^{0c} + u^0 u^c \right)
        \left( \psi_{,c} \psi^* - \psi^*_{,c} \psi \right) \bigg]
   \nonumber \\
   & & \qquad
       + {\hbar^2 \over m}
       \left[ \left( \psi_{,(a} + \psi_{,c} u^c u_{(a} \right)
       \left( \psi^*_{,b)} + \psi^*_{,d} u^d u_{b)} \right)
       - {1 \over 3} \left( g_{ab} + u_a u_b \right)
       \left( \psi^{;c} \psi^*_{,c}
       + |\psi_{,c} u^c|^2 \right) \right].
   \label{fluid-Schrodinger}
\eea
The energy-frame condition, $q_a \equiv 0$, gives
\bea
   & & m |\psi|^2 c^2 u^0 \left( u^0 u_a + c t_{,a} \right)
       = - i \hbar c \left[
       {1 \over 2} \left( \psi_{,a} \psi^*
       - \psi^*_{,a} \psi \right) u^0
       + \left( \psi_{,c} \psi^* - \psi^*_{,c} \psi \right) u^c
       \left( u^0 u_a + {1 \over 2} c t_{,a} \right)
       \right]
   \nonumber \\
   & & \qquad
       - {\hbar^2 \over m} \left[
       |\psi_{,c} u^c |^2 u_a
       + {1 \over 2} \left( \psi_{,a} \psi^*_{,c}
       + \psi^*_{,a} \psi_{,c} \right) u^c \right].
   \label{E-frame-1}
\eea
Contracting with $c t^{;a} = g^{ab} c t_{,b} = g^{0a}$, we have
\bea
   & & m |\psi|^2 c^2 u^0 \left( u^0 u^0 + g^{00} \right)
       = - i \hbar c
       \left[ {1 \over 2} \left( g^{0c} u^0
       + g^{00} u^c \right) + u^0 u^0 u^c \right]
       \left( \psi_{,c} \psi^* - \psi^*_{,c} \psi \right)
   \nonumber \\
   & & \qquad
       - {\hbar^2 \over m} \left[
       |\psi_{,c} u^c |^2 u^0
       + {1 \over 2} g^{0c} \left( \psi_{,c} \psi^*_{,d}
       + \psi^*_{,c} \psi_{,d} \right) u^d \right].
   \label{E-frame-2}
\eea
We have $c t_{,a} = \delta^0_a$, thus $u^0 u_a + c t_{,a} = h^0_a$ and $u^0 u^0 + g^{00} = h^{00}$ are components of the spatial projection-tensor. The Schr\"odinger equation (\ref{relativistic-Schrodinger-eq}) gives
\bea
   & & m |\psi|^2 c^2 ( g^{00} + 1 )
       = - i \hbar c g^{0c}
       \left( \psi_{,c} \psi^* - \psi^*_{,c} \psi \right)
       + {\hbar^2 \over 2 m} \left[
       ( \Box \psi ) \psi^* + ( \Box \psi^* ) \psi
       - 16 \pi \ell_s | \psi |^4 \right].
\eea
Using these the fluid quantities become
\bea
   & & \mu = m |\psi|^2 c^2
       + i \hbar {1 \over 2} {c \over u^0} \left( g^{0c} u^0
       - g^{00} u^c \right)
       \left( \psi_{,c} \psi^* - \psi^*_{,c} \psi \right)
   \nonumber \\
   & & \qquad
       + {\hbar^2 \over 2 m} \left\{
       \psi^{;c} \psi^*_{,c}
       - g^{0c}
       \left( \psi_{,c} \psi^*_{,d}
       + \psi^*_{,c} \psi_{,d} \right) {u^d \over u^0}
       - {1 \over 2} \left[ ( \Box \psi ) \psi^*
       + ( \Box \psi^* ) \psi \right]
       + 12 \pi \ell_s | \psi |^4
       \right\},
   \nonumber \\
   & & p = i \hbar {1 \over 6} {c \over u^0}
       \left( g^{0c} u^0 - g^{00} u^c \right)
       \left( \psi_{,c} \psi^* - \psi^*_{,c} \psi \right)
   \nonumber \\
   & & \qquad
       - {\hbar^2 \over 6 m} \left\{
       \psi^{;c} \psi^*_{,c}
       + g^{0c}
       \left( \psi_{,c} \psi^*_{,d}
       + \psi^*_{,c} \psi_{,d} \right) {u^d \over u^0}
       + {3 \over 2} \left[ ( \Box \psi ) \psi^*
       + ( \Box \psi^* ) \psi \right]
       - 12 \pi \ell_s | \psi |^4
       \right\}.
\eea
As we have
\bea
   & & g^{0c} u^0 - g^{00} u^c
       = (g^{0c} + u^0 u^c ) u^0
       - ( g^{00} + u^0 u^0 ) u^c, \quad {\rm etc.},
\eea
using Equations (\ref{E-frame-1}) and (\ref{E-frame-2}), we finally have
\bea
   & & \mu = m |\psi|^2 c^2
       + {\hbar^2 \over 4 m} \left\{
       2 \psi^{;c} \psi^*_{,c}
       - 2 g^{0c} \left( \psi_{,c} \psi^*_{,d}
       + \psi^*_{,c} \psi_{,d} \right)
       {u^d \over u^0}
       - \left[ ( \Box \psi ) \psi^*
       + ( \Box \psi^* ) \psi \right]
       + 24 \pi \ell_s | \psi |^4 \right\}
   \nonumber \\
   & & \qquad
       + {\hbar^2 \over 4 m} {1 \over |\psi|^2}
       \bigg\{
       \left( \psi^{;c} \psi^* - \psi^{*;c} \psi \right)
       \left( \psi_{,c} \psi^* - \psi^*_{,c} \psi \right)
       - g^{00} \bigg[
       \left( \psi_{,c} \psi^* - \psi^*_{,c} \psi \right) {u^c \over u^0} \bigg]^2
       \bigg\}
   \nonumber \\
   & & \qquad
       + {i \hbar^3 \over 4 m^2} {1 \over c |\psi|^2}
       \left[
       - \left( \psi^{;c} \psi^* - \psi^{*;c} \psi \right)
       \left( \psi_{,c} \psi^*_{,d}
       + \psi^*_{,c} \psi_{,d} \right)
       {u^d \over u^0}
       + g^{0c}
       \left( \psi_{,c} \psi^*_{,d}
       + \psi^*_{,c} \psi_{,d} \right)
       {u^d \over u^0}
       \left( \psi_{,e} \psi^* - \psi^*_{,e} \psi \right)
       {u^e \over u^0}
       \right],
   \nonumber \\
   & & p = - {\hbar^2 \over 6 m} \left\{
       \psi^{;c} \psi^*_{,c}
       + g^{0c} \left( \psi_{,c} \psi^*_{,d}
       + \psi^*_{,c} \psi_{,d} \right)
       {u^d \over u^0}
       + {3 \over 2} \left[ ( \Box \psi ) \psi^*
       + ( \Box \psi^* ) \psi \right]
       - 12 \pi \ell_s | \psi |^4 \right\}
   \nonumber \\
   & & \qquad
       + {\hbar^2 \over 12 m} {1 \over |\psi|^2}
       \bigg\{
       \left( \psi^{;c} \psi^* - \psi^{*;c} \psi \right)
       \left( \psi_{,c} \psi^* - \psi^*_{,c} \psi \right)
       - g^{00}
       \bigg[
       \left( \psi_{,c} \psi^* - \psi^*_{,c} \psi \right)
       {u^c \over u^0} \bigg]^2
       \bigg\}
   \nonumber \\
   & & \qquad
       + {i \hbar^3 \over 12 m^2} {1 \over c |\psi|^2}
       \bigg[
       - \left( \psi^{;c} \psi^* - \psi^{*;c} \psi \right)
       \left( \psi_{,c} \psi^*_{,d}
       + \psi^*_{,c} \psi_{,d} \right)
       {u^d \over u^0}
       + g^{0c}
       \left( \psi_{,c} \psi^*_{,d}
       + \psi^*_{,c} \psi_{,d} \right)
       {u^d \over u^0}
       \left( \psi_{,e} \psi^* - \psi^*_{,e} \psi \right)
       {u^e \over u^0}
       \bigg],
%   \nonumber \\
   \nonumber
\eea
\bea
   & & \pi_{ab} = {\hbar^2 \over m} \left[
       \left( \psi_{,(a} + \psi_{,c} u^c u_{(a} \right)
       \left( \psi^*_{,b)} + \psi^*_{,c} u^c u_{b)} \right)
       - {1 \over 3} \left( g_{ab} + u_a u_b \right)
       \left( \psi^{;c} \psi^*_{,c}
       + |\psi_{,c} u^c|^2 \right) \right]
   \nonumber \\
   & & \qquad
       + {\hbar^2 \over m} {1 \over |\psi|^2}
       \bigg\{ {1 \over 4}
       \left( \psi_{,a} \psi^* - \psi^*_{,a} \psi \right)
       \left( \psi_{,b} \psi^* - \psi^*_{,b} \psi \right)
       + {1 \over 2}
       \left( \psi_{,(a} \psi^* - \psi^*_{,(a} \psi \right)
       u_{b)}
       \left( \psi_{,c} \psi^* - \psi^*_{,c} \psi \right) u^c
   \nonumber \\
   & & \qquad \qquad
       - {1 \over 2} \left( u^0 u_{(a} \delta^0_{b)}
       + {1 \over 2} \delta^0_a \delta^0_b \right)
       \left[
       \left( \psi_{,c} \psi^* - \psi^*_{,c} \psi \right)
       {u^c \over u^0} \right]^2
   \nonumber \\
   & & \qquad \qquad
       - {1 \over 3} \left( g_{ab} + u_a u_b \right)
       {1 \over 4}
       \bigg[
       \left( \psi^{;c} \psi^* - \psi^{*;c} \psi \right)
       \left( \psi_{,c} \psi^* - \psi^*_{,c} \psi \right)
       - g^{00} \left[
       \left( \psi_{,c} \psi^* - \psi^*_{,c} \psi \right)
       {u^c \over u^0} \right]^2
       \bigg]
       \bigg\}
   \nonumber \\
   & & \qquad
       + {i \hbar^3 \over m^2} {1 \over c (u^0)^2 |\psi|^2}
       \bigg\{
       \left( u^0 u_{(a} + \delta^0_{(a} \right) u_{b)}
       |\psi_{,c} u^c|^2
       + {1 \over 2} \left( u^0 u_{(a} + \delta^0_{(a} \right)
       \left( \psi_{,b)} \psi^*_{,c} + \psi^*_{,b)} \psi_{,c} \right) u^c
   \nonumber \\
   & & \qquad \qquad
       - {1 \over 3} \left( g_{ab} + u_a u_b \right)
       \left[ {1 \over 2} g^{0c}
       \left( \psi_{,c} \psi^*_{,d} + \psi^*_{,c} \psi_{,d} \right) u^d
       + |\psi_{,c} u^c|^2 u^0 \right]
       \bigg\}
       \left( \psi_{,e} \psi^* - \psi^*_{,e} \psi \right) u^e
   \nonumber \\
   & & \qquad
       + {\hbar^4 \over m^3} {1 \over c^2 (u^0)^2 |\psi|^2}
       \bigg\{
       \left[ |\psi_{,c} u^c|^2 u_a
       + {1 \over 2}
       \left( \psi_{,a} \psi^*_{,c} + \psi^*_{,a} \psi_{,c} \right) u^c \right]
       \left[ |\psi_{,c} u^c|^2 u_b
       + {1 \over 2}
       \left( \psi_{,b} \psi^*_{,c} + \psi^*_{,b} \psi_{,c} \right) u^c \right]
   \nonumber \\
   & & \qquad \qquad
       - {1 \over 3} \left( g_{ab} + u_a u_b \right)
       \left[ |\psi_{,e} u^e|^4
       + {1 \over 4}
       \left( \psi_{,c} \psi^*_{,d} + \psi^*_{,c} \psi_{,d} \right) u^c
       \left( \psi^{;d} \psi^*_{,e} + \psi^{*;d} \psi_{,e} \right) u^e \right]
       \bigg\}.
   \label{fluid-K}
\eea
For ${\hbar^2 \over m^2 c^2} \rightarrow 0$, we have
\bea
   & & \mu = m |\psi|^2 c^2, \quad p = 0 = \pi_{ab},
\eea
thus behave as a zero-pressure fluid. Only for ${\hbar^2 \over m^2 c^2} \rightarrow 0$, $m |\psi|^2$ can be properly identified as the fluid density $\varrho (\equiv \mu/c^2)$.

%%%%%%%%%%%%%%%%%%%%%%%%%%%%%%%%%%%%%%%%%%%%%%%%%%%%%%%%%%%%%%%%%
\subsection{Relativistic Madelung formulation}
                                 \label{sec:M-transformation-cov}

The Madelung transformation is in Eq.\ (\ref{M-transformation}). Applying the Madelung transformation to Eq.\ (\ref{relativistic-Schrodinger-eq}), we have
\bea
   & & {\Box \sqrt{\varrho} \over \sqrt{\varrho}}
       - {8 \pi \ell_s \over m} \varrho
       - {m^2 \over \hbar^2} \left( u^{;c} u_{,c} + c^2
       + c^2 g^{00} - 2 c g^{0c} u_{,c} \right)
       + {i m \over \hbar} {1 \over \varrho}
       \left[ \left( \varrho u^{;c} \right)_{;c}
       - c g^{0c} \varrho_{,c}
       + c g^{ab} \Gamma^0_{ab} \varrho \right]
       = 0.
   \label{fluid-Klein}
\eea
The imaginary and real parts, respectively, give
\bea
   & & \left( \varrho u^{;c} \right)_{;c}
       = c \varrho^{;0}
       - c g^{ab} \Gamma^0_{ab} \varrho,
   \label{imaginary-Klein} \\
   & & u^{;c} u_{,c} + c^2
       - {\hbar^2 \over m^2} \left(
       {\Box \sqrt{\varrho} \over \sqrt{\varrho}}
       - {8 \pi \ell_s \over m} \varrho \right)
       = - c^2 g^{00} + 2 c u^{;0}.
   \label{real-Klein}
\eea

Using the Madelung transformation in Eq.\ (\ref{M-transformation}) on Eq.\ (\ref{fluid-Schrodinger}), we have
\bea
   & & \mu = \varrho
       \left( c u^0 - u_{,c} u^c \right)^2
       + {\hbar^2 \over m^2} \left[
       {1 \over 2} \sqrt{\varrho}^{;c} \sqrt{\varrho}_{,c}
       + ( \sqrt{\varrho}_{,c} u^c )^2
       + {1 \over 2} \sqrt{\varrho} \Box \sqrt{\varrho}
       - {2 \pi \ell_s \over m} \varrho^2
       \right],
   \nonumber \\
   & & p ={1 \over 3}  \varrho \left[
       \left( c u^0 - u_{,c} u^c \right)^2 - c^2 \right]
       + {\hbar^2 \over 3 m^2} \left[
       - {1 \over 2} \sqrt{\varrho}^{;c} \sqrt{\varrho}_{,c}
       + ( \sqrt{\varrho}_{,c} u^c )^2
       - {1 \over 2} \sqrt{\varrho} \Box \sqrt{\varrho}
       - {2 \pi \ell_s \over m} \varrho^2
       \right],
   \nonumber \\
   & & q_a = \varrho \left( c u^0 - u_{,c} u^c \right)
       \left[ \left( u - c^2 t \right)_{,a}
       - \left( c u^0 - u_{,c} u^c \right) u_a \right]
       - {\hbar^2 \over m^2} ( \sqrt{\varrho}_{,c} u^c ) \left[
       \sqrt{\varrho}_{,a}
       + ( \sqrt{\varrho}_{,d} u^d ) u_a \right],
   \nonumber \\
   & & \pi_{ab} = \varrho \bigg\{
       \left[ \left( u - c^2 t \right)_{,a}
       - \left( c u^0 - u_{,c} u^c \right) u_{a} \right]
       \left[ \left( u - c^2 t \right)_{,b}
       - \left( c u^0 - u_{,c} u^c \right) u_{b} \right]
   \nonumber \\
   & & \qquad \qquad
       - {1 \over 3} \left( g_{ab} + u_a u_b \right)
       \left[ u^{;c} u_{,c}
       - 2 c g^{0c} u_{,c} + c^2 g^{00}
       + \left( c u^0 - u_{,c} u^c \right)^2
       \right] \bigg\}
   \nonumber \\
   & & \qquad
       + {\hbar^2 \over m^2} \left\{
       \left[ \sqrt{\varrho}_{,a}
       + ( \sqrt{\varrho}_{,c} u^c ) u_a \right]
       \left[ \sqrt{\varrho}_{,b}
       + ( \sqrt{\varrho}_{,d} u^d ) u_b \right]
       - {1 \over 3} \left( g_{ab} + u_a u_b \right)
       \left[ \sqrt{\varrho}^{;c} \sqrt{\varrho}_{,c}
       + ( \sqrt{\varrho}_{,c} u^c )^2 \right]
       \right\},
   \label{fluid-Klein-Madelung}
\eea
where we used Equation (\ref{real-Klein}).

The energy-frame condition, $q_a \equiv 0$, gives
\bea
   & & u_{,a} = c^2 t_{,a}
       + \left( c u^0 - u_{,c} u^c \right) u_a
       + {\hbar^2 \over m^2} {\sqrt{\varrho}_{,c} u^c
       \over \varrho ( c u^0 - u_{,c} u^c )} \left[
       \sqrt{\varrho}_{,a}
       + ( \sqrt{\varrho}_{,d} u^d ) u_a \right].
   \label{energy-frame-KM}
\eea
Using this, Eq.\ (\ref{real-Klein}) becomes
\bea
   & & \left( c u^0 - u_{,c} u^c \right)^2
       = c^2
       - {\hbar^2 \over m^2}
       \left( {\Box \sqrt{\varrho} \over \sqrt{\varrho}}
       - {8 \pi \ell_s \over m} \varrho \right)
       + {\hbar^4 \over m^4} {(\sqrt{\varrho}_{,c} u^c)^2
       \over \varrho^2 ( c u^0 - u_{,c} u^c )^2}
       g^{ab} \left[ \sqrt{\varrho}_{,a}
       + ( \sqrt{\varrho}_{,d} u^d ) u_a \right]
       \left[ \sqrt{\varrho}_{,b}
       + ( \sqrt{\varrho}_{,d} u^d ) u_b \right].
   \nonumber \\
\eea
Thus, Eq.\ (\ref{fluid-Klein-Madelung}) gives
\bea
   & & \mu =
       \varrho c^2
       + {\hbar^2 \over m^2} \left[
       {1 \over 2} \sqrt{\varrho}^{;c} \sqrt{\varrho}_{,c}
       + ( \sqrt{\varrho}_{,c} u^c )^2
       - {1 \over 2} \sqrt{\varrho} \Box \sqrt{\varrho}
       + {6 \pi \ell_s \over m} \varrho^2
       \right]
   \nonumber \\
   & & \qquad
       + {\hbar^4 \over m^4} {(\sqrt{\varrho}_{,c} u^c)^2
       \over \varrho ( c u^0 - u_{,c} u^c )^2}
       g^{ab} \left[ \sqrt{\varrho}_{,a}
       + ( \sqrt{\varrho}_{,d} u^d ) u_a \right]
       \left[ \sqrt{\varrho}_{,b}
       + ( \sqrt{\varrho}_{,d} u^d ) u_b \right],
   \nonumber \\
   & & p = {\hbar^2 \over 3 m^2} \left[
       - {1 \over 2} \sqrt{\varrho}^{;c} \sqrt{\varrho}_{,c}
       + ( \sqrt{\varrho}_{,c} u^c )^2
       - {3 \over 2} \sqrt{\varrho} \Box \sqrt{\varrho}
       + {6 \pi \ell_s \over m} \varrho^2
       \right]
   \nonumber \\
   & & \qquad
       + {\hbar^4 \over 3 m^4} {(\sqrt{\varrho}_{,c} u^c)^2
       \over \varrho ( c u^0 - u_{,c} u^c )^2}
       g^{ab} \left[ \sqrt{\varrho}_{,a}
       + ( \sqrt{\varrho}_{,d} u^d ) u_a \right]
       \left[ \sqrt{\varrho}_{,b}
       + ( \sqrt{\varrho}_{,d} u^d ) u_b \right],
   \nonumber \\
   & & \pi_{ab} = {\hbar^2 \over m^2} \left\{
       \left[ \sqrt{\varrho}_{,a}
       + ( \sqrt{\varrho}_{,c} u^c ) u_a \right]
       \left[ \sqrt{\varrho}_{,b}
       + ( \sqrt{\varrho}_{,d} u^d ) u_b \right]
       - {1 \over 3} \left( g_{ab} + u_a u_b \right)
       \left[
       \sqrt{\varrho}^{;c} \sqrt{\varrho}_{,c}
       + ( \sqrt{\varrho}_{,c} u^c )^2 \right] \right\}
   \nonumber \\
   & & \qquad
       + {\hbar^4 \over m^4} {(\sqrt{\varrho}_{,c} u^c)^2
       \over \varrho ( c u^0 - u_{,c} u^c )^2}
       \bigg\{ \left[ \sqrt{\varrho}_{,a}
       + ( \sqrt{\varrho}_{,d} u^d ) u_a \right]
       \left[ \sqrt{\varrho}_{,b}
       + ( \sqrt{\varrho}_{,d} u^d ) u_b \right]
   \nonumber \\
   & & \qquad \qquad
       - {1 \over 3} \left( g_{ab} + u_a u_b \right)
       g^{cd} \left[ \sqrt{\varrho}_{,c}
       + ( \sqrt{\varrho}_{,e} u^e ) u_c \right]
       \left[ \sqrt{\varrho}_{,d}
       + ( \sqrt{\varrho}_{,e} u^e ) u_d \right]
       \bigg\}.
   \label{fluid-KM}
\eea
%These also follow from Eq.\ (\ref{fluid-K}).
For ${\hbar^2 \over m^2 c^2} \rightarrow 0$, we have
\bea
   & & \mu = \varrho c^2, \quad p = 0 = \pi_{ab},
\eea
thus behave as a zero-pressure fluid; only for ${\hbar^2 \over m^2 c^2} \rightarrow 0$, $\varrho$ can be properly identified as the fluid density.

%%%%%%%%%%%%%%%%%%%%%%%%%%%%%%%%%%%%%%%%%%%%%%%%%%%%%%%%%%%%%%%%%
%
%
%
%%%%%%%%%%%%%%%%%%%%%%%%%%%%%%%%%%%%%%%%%%%%%%%%%%%%%%%%%%%%%%%%%
\section{Cosmological 1PN approximation}
                                      \label{sec:1PN-approximation}

%%%%%%%%%%%%%%%%%%%%%%%%%%%%%%%%%%%%%%%%%%%%%%%%%%%%%%%%%%%%%%%%%
\subsection{Curvature and fluid quantities}

Our metric convention is
\bea
   & & g_{00} = - \left( 1 + 2 \alpha \right), \quad
       g_{0i} = - \chi_i, \quad
       g_{ij} = a^2 \left( 1 + 2 \varphi \right) \delta_{ij},
   \label{metric-pert}
\eea
with
\bea
   & & \alpha \equiv {\Phi \over c^2}, \quad
       \varphi \equiv - {\Psi \over c^2}, \quad
       \chi_i \equiv a {P_i \over c^3}.
\eea
We consider the flat cosmological background, and index $0 = c t$. The spatial indices of $\chi_i$ and $P_i$ are raised and lowered using $\delta_{ij}$ and its inverse. In order to properly include the 1PN expansion, we have to consider $c^{-4}$-order in $g_{00}$, see Eq.\ (\ref{Chandrasekhar-notation}); thus, $\Phi$ includes $c^{-2}$ order, and we expand the inverse metric $g^{00}$ to $\Phi^2$ order. Here are the metric tensor, connection and curvatures to the 1PN order in PN expansion
\bea
   & & g_{00} = - \left( 1 + 2 {\Phi \over c^2} \right), \quad
       g_{0i} = - a {P_i \over c^3}, \quad
       g_{ij} = a^2 \left( 1 - 2 {\Psi \over c^2} \right) \delta_{ij},
   \nonumber \\
   & & g^{00} = - \left( 1 - 2 {\Phi \over c^2}
       + 4 {\Phi^2 \over c^4} \right), \quad
       g^{0i} = - {1 \over a} {P^i \over c^3}, \quad
       g^{ij} = {1 \over a^2}
       \left( 1 + 2 {\Psi \over c^2} \right) \delta^{ij},
   \\
   & & \Gamma^0_{00} = {\dot \Phi \over c^3}
       - {1 \over c^5} \left( 2 \Phi \dot \Phi
       + {1 \over a} P^i \Phi_{,i} \right), \quad
       \Gamma^0_{0i} = {\Phi_{,i} \over c^2}
       - {1 \over c^4} \left( 2 \Phi \Phi_{,i}
       + a H P_i \right),
   \nonumber \\
   & &
       \Gamma^0_{ij} = {1 \over c} a^2 H \delta_{ij}
       - {1 \over c^3} a^2 \left[ \dot \Psi
       + 2 H \left( \Phi + \Psi \right) \right] \delta_{ij}
       + {1 \over c^3} a P_{(i,j)}, \quad
       \Gamma^i_{00} = {1 \over c^2} {1 \over a^2} \Phi^{,i}
       + {1 \over c^4} {1 \over a^2}
       \left[ 2 \Psi \Phi^{,i}
       - ( a P^i )^{\displaystyle{\cdot}} \right],
   \nonumber \\
   & &
       \Gamma^i_{0j} = {1 \over c} H \delta^i_j
       - {\dot \Psi \over c^3} \delta^i_j
       + {1 \over c^3} {1 \over 2 a} \left( P_j^{\;\;,i} - P^i_{\;\;,j} \right), \quad
       \Gamma^i_{jk} = - {1 \over c^2} \left( \Psi_{,k} \delta^i_j
       + \Psi_{,j} \delta^i_k - \Psi^{,i} \delta_{jk} \right).
\eea
The Riemann curvature is
\bea
   & & R^0_{\;\;00i}
       = - {1 \over c^5} \left( \ddot a P_i
       - {1 \over a} \Phi_{,ij} P^j \right), \quad
       R^0_{\;\;0ij}
       = L^{-2} {\cal O} (c^{-6}),
   \nonumber \\
   & & R^0_{\;\;i0j}
       = {1 \over c^2} \left( a \ddot a \delta_{ij}
       - \Phi_{,ij} \right)
       + {1 \over c^4} \bigg\{
       - a^2 \left[ \dot \Psi + 2 H \left( \Phi + \Psi \right)
       \right]^{\displaystyle{\cdot}} \delta_{ij}
       - a^2 H \left[ - \dot \Phi + 2 H \left( \Phi + \Psi \right)
       \right] \delta_{ij}
   \nonumber \\
   & & \qquad
       + \Phi_{,i} \Phi_{,j} + 2 \Phi \Phi_{,ij}
       - \Psi_{,i} \Phi_{,j} - \Psi_{,j} \Phi_{,i}
       + \Psi^{,k} \Phi_{,k} \delta_{ij}
       + \left( a P_{(i,j)} \right)^{\displaystyle{\cdot}} \bigg\},
   \nonumber \\
   & & R^0_{\;\;ijk}
       = - {1 \over c^3} \left[ 2 a^2 ( \dot \Psi
       + H \Phi )_{,[j} \delta_{k]i}
       + a P_{[j,k]i} \right],
   \nonumber \\
   & & R^i_{\;\;00j}
       = {1 \over c^2} \left( {\ddot a \over a} \delta^i_j
       - {1 \over a^2} \Phi^{,i}_{\;\;j} \right)
       + {1 \over c^4} \bigg\{
       - \left[ \ddot \Psi + H \left( \dot \Phi + 2 \dot \Psi \right) \right] \delta^i_j
   \nonumber \\
   & & \qquad
       + {1 \over a^2} \left( - 2 \Psi \Phi^{,i}_{\;\;j}
       + \Phi^{,i} \Phi_{,j}
       - \Psi^{,i} \Phi_{,j} - \Psi_{,j} \Phi^{,i}
       + \Psi^{,k} \Phi_{,k} \delta^i_j \right)
       + {1 \over 2 a^2} \left[
       a \left( P^i_{\;\;,j} + P_j^{\;\;,i} \right)
       \right]^{\displaystyle{\cdot}} \bigg\},
   \nonumber \\
   & & R^i_{\;\;0jk}
       = - {1 \over c^3} \left[
       2 ( \dot \Psi + H \Phi )_{,[j} \delta^i_{k]}
       + {1 \over a} P_{[j,k]}^{\;\;\;\;\;\;i} \right],
   \nonumber \\
   & & R^i_{\;\;j0k}
       = {1 \over c^3} \left[
       ( \dot \Psi + H \Phi )^{,i} \delta_{jk}
       - ( \dot \Psi + H \Phi )_{,j} \delta^i_k
       + {1 \over 2 a} ( P^i_{\;\;,j} - P_j^{\;\;,i} )_{,k} \right],
   \nonumber \\
   & & R^i_{\;\;jk\ell}
       = {2 \over c^2} \left(
       a^2 H^2 \delta^i_{[k} \delta_{\ell]j}
       + \Psi^{,i}_{\;\;[k} \delta_{\ell]j}
       - \Psi_{,j[k} \delta^i_{\ell]} \right).
\eea
The Ricci and the scalar curvature are
\bea
   & & R^0_0
       = {1 \over c^2} \left( 3 {\ddot a \over a}
       - {\Delta \over a^2} \Phi \right)
       - {1 \over c^4} \bigg\{
       3 \left[ \ddot \Psi + H \left( \dot \Phi + 2 \dot \Psi \right) + 2 {\ddot a \over a} \Phi \right]
       - {1 \over a^2} \left[ 2 \left( \Phi - \Psi \right) \Delta \Phi
       + \left( \Phi + \Psi \right)^{,i} \Phi_{,i} \right]
       - {1 \over a^2} \left( a P^i_{\;\;,i} \right)^{\displaystyle{\cdot}} \bigg\},
   \nonumber \\
   & & R^0_i
       = {1 \over c^3} \left[ - 2 \left( \dot \Psi + H \Phi
       \right)_{,i}
       + {1 \over 2 a} \left( P^k_{\;\;,ki}
       - \Delta P_i \right) \right], \quad
%   \nonumber \\
%   & & R^i_0
%       = {1 \over c^3} {1 \over a^2} \left[ 2 \left( \dot \Psi + H \Phi \right)^{,i} - {1 \over 2 a} \left( P^{k\;\;\;i}_{\;\;,k} - \Delta P^i \right) \right],
   R^i_j
       = {1 \over c^2} \left[ \left( {\ddot a \over a}
       + 2 H^2 \right) \delta^i_j
       + {\Delta \over a^2} \Psi \delta^i_j
       + {1 \over a^2} \left( \Psi - \Phi \right)^{,i}_{\;\;j}
       \right],
   \nonumber \\
   & & R
       = {1 \over c^2} \left[ 6 \left( {\ddot a \over a}
       + H^2 \right)
       + 2 {\Delta \over a^2} \left( 2 \Psi - \Phi \right) \right].
\eea
Using Chandrasekhar's 1PN notation in Eq.\ (\ref{Chandrasekhar-notation}) we recover Equations (1)-(8) in \cite{Hwang-Noh-Puetzfeld-2008}.

The normalized fluid four-vector is introduced as
\bea
   & & u_i \equiv a \gamma {v_i \over c}
       = a \left( 1 + {v^2 \over 2 c^2} \right) {v_i \over c}, \quad
       u_0 = - 1 -  {1 \over c^2}
       \left( {1 \over 2} v^2 + \Phi \right)
       - {1 \over c^4} \left[
       \left( {3 \over 8} v^2 + {1 \over 2} \Phi + \Psi \right) v^2
       - {1 \over 2} \Phi^2 + P^i v_i \right],
   \nonumber \\
   & &
       u^i = {1 \over a c} \left[ \left( 1 + {v^2 \over 2 c^2} + 2 {\Psi \over c^2} \right) v^i + {P^i \over c^2} \right], \quad
       u^0 = 1 + {1 \over c^2}
       \left( {1 \over 2} v^2 - \Phi \right)
       + {1 \over c^4} \left[
       \left( {3 \over 8} v^2 - {1 \over 2} \Phi + \Psi \right) v^2
       + {3 \over 2} \Phi^2 \right],
   \label{four-vector-1PN}
\eea
where $v^2 \equiv v^i v_i$.

We introduce
\bea
   & & \mu \equiv \varrho
       \left( c^2 + \Pi \right), \quad
       q_i \equiv {a \over c} Q_i, \quad
       \pi_{ij} \equiv a^2 \Pi_{ij},
   \label{fluid-PN}
\eea
where $\varrho$ and $\varrho \Pi$ are mass density and internal energy density, respectively. The indices of $Q_i$ and $\Pi_{ij}$ are raised and lowered using $\delta_{ij}$ and its inverse;
we have
\bea
   & & q_i \equiv {a \over c} Q_i, \quad
       q_0 = - {1 \over c^2} Q_i v^i, \quad
       q^i = {1 \over ac} Q^i, \quad
       q^0 = { 1\over c^2} Q_i v^i,
   \nonumber \\
   & & \pi_{ij} \equiv a^2 \Pi_{ij}, \quad
       \pi_{0i} = - {a \over c} \Pi_{ij} v^j, \quad
       \pi_{00} = {1 \over c^2} \Pi_{ij} v^i v^j,
   \nonumber \\
   & & \pi^i_j = \left( 1 + 2 {\Psi \over c^2} \right)
       \Pi^i_j, \quad
       \pi^i_0 = - {1 \over c a} \Pi^i_j v^j, \quad
       \pi^0_i = {a \over c} \Pi_{ij} v^j, \quad
       \pi^0_0 = - {1 \over c^2} \Pi_{ij} v^i v^j,
\eea
thus, $\pi^c_c = 0$ implies
\bea
   & & \Pi^i_i = {1 \over c^2} \Pi_{ij} v^i v^j.
\eea
The energy-momentum tensor gives
\bea
   & & T^0_0 = - \varrho c^2 - \varrho \left( \Pi + v^2 \right)
       - {1 \over c^2} \left[
       \varrho \left( \Pi + v^2 + 2 \Psi \right) v^2
       + \varrho P^i v_i
       + p v^2 + 2 Q_i v^i
       + \Pi_{ij} v^i v^j \right],
   \nonumber \\
   & & T^0_i = \varrho c a v_i
       + {a \over c} \left[ \varrho v_i
       \left( \Pi + v^2 - \Phi \right)
       + p v_i + Q_i + \Pi_{ij} v^j \right],
   \nonumber \\
   & & T^i_j = \varrho v^i v_j + p \delta^i_j + \Pi^i_j
       + {1 \over c^2} \left[ \varrho v^i v_j
       \left( \Pi + v^2 + 2 \Psi \right)
       + \varrho P^i v_j + p v^i v_j
       + Q^i v_j + Q_j v^i + 2 \Psi \Pi^i_j \right],
   \label{Tab-fluid-1PN}
\eea
thus $T = - \varrho c^2 - \varrho \Pi + 3 p$.

%%%%%%%%%%%%%%%%%%%%%%%%%%%%%%%%%%%%%%%%%%%%%%%%%%%%%%%%%%%%%%%%%
\subsection{1PN hydrodynamic formulation}

We consider Einstein's equation in a form
\bea
   & & R^a_b = {8 \pi G \over c^4}
       \left( T^a_b - {1 \over 2} T \delta^a_b \right)
       + \Lambda \delta^a_b.
\eea
The $R^0_0$, $R^0_i$ and $R^i_j$ components, respectively, give
\bea
   & & {1 \over c^2} \left( {\Delta \over a^2} \Phi
       - 4 \pi G \varrho - 3 {\ddot a \over a}
       + \Lambda c^2 \right)
       + {1 \over c^4} \bigg\{
       3 \left[ \ddot \Psi + H ( \dot \Phi + 2 \dot \Psi )
       + 2 {\ddot a \over a} \Phi \right]
       - {1 \over a^2} \left[ 2 ( \Phi - \Psi ) \Delta \Phi
       + ( \Phi + \Psi )^{,i} \Phi_{,i} \right]
   \nonumber \\
   & & \qquad
       - {1 \over a^2} ( a P^i_{\;\;,i} )^{\displaystyle{\cdot}}
       - 8 \pi G \left[ \varrho \left( {1 \over 2} \Pi
       + v^2 \right)
       + {3 \over 2} p \right] \bigg\} = 0,
   \label{PN-1} \\
   & & {1 \over c^3} \left[ - ( \dot \Psi + H \Phi )_{,i}
       + {1 \over 4 a} ( P^k_{\;\;,ki} - \Delta P_i )
       - 4 \pi G \varrho a v_i \right] = 0,
   \label{PN-2} \\
   & & {1 \over c^2} \left[ \left( {\ddot a \over a} + 2 H^2
       \right) \delta^i_j
       + {\Delta \over a^2} \Psi \delta^i_j
       + {1 \over a^2} ( \Psi - \Phi )^{,i}_{\;\; j} \right]
       = {8 \pi G \over c^4} \left[ \varrho v^i v_j + \Pi^i_j
       + {1 \over 2} ( \varrho c^2 + \varrho \Pi
       - p ) \delta^i_j \right]
       + \Lambda \delta^i_j,
   \label{PN-3}
\eea
where the left-hand sides are derived up to 1PN order; we kept $c^{-4}$-order in Eq.\ (\ref{PN-3}) only to have proper cosmological background equation. In the cosmological background, to the background order, from Eqs.\ (\ref{PN-1}) and (\ref{PN-3}), we have
\bea
   & & {\ddot a \over a} = - {4 \pi G \over 3} \left[
       \varrho_b \left( 1 + {\Pi_b \over c^2} \right)
       + 3 {p_b \over c^2} \right]
       + {\Lambda c^2 \over 3}, \quad
       {\dot a^2 \over a^2} = {8 \pi G \over 3} \varrho_b
       \left( 1 + {\Pi_b \over c^2} \right)
       + {\Lambda c^2 \over 3}.
   \label{BG-eqs}
\eea
Subtracting the background equation, Eq.\ (\ref{PN-3}) becomes
\bea
   & & {\Delta \over a^2} \Psi \delta^i_j
       + {1 \over a^2} ( \Psi - \Phi )^{,i}_{\;\; j}
       = 4 \pi G \delta \varrho,
\eea
Thus, trace and tracefree parts, respectively, give
\bea
   & & {\Delta \over a^2} \Psi = 4 \pi G \delta \varrho, \quad
       \Psi = \Phi.
\eea
As mentioned, $\Phi$ still includes 1PN order contribution besides the 0PN (Newtonian) one. Compared with Chandrasekhar's notation
\bea
   & & \Phi = - U - {1 \over c^2}
       \left( 2 \Upsilon - U^2 \right), \quad
       \Psi = - V, \quad
       v_i = \overline v_i - {1 \over c^2} \left[
       \left( \Phi + 2 \Psi \right) \overline v_i + P_i \right],
   \label{Chandrasekhar-notation}
\eea
where
\bea
   & & {1 \over a c} \overline v^i
       \equiv {d x^i \over d x^0}
       = {u^i \over u^0}, \quad
       u_i \equiv {a \over c} \gamma v_i.
\eea
Using $U$ and $V$, we have $V = U$ to 0PN order; $\overline v_i$ is used in \cite{Chandrasekhar-1965, Hwang-Noh-Puetzfeld-2008}. Subtracting the background equation, Eqs.\ (\ref{PN-1}) and (\ref{PN-2}) give
\bea
   & & {\Delta \over a^2} U
       + 4 \pi G ( \varrho - \varrho_b )
       + {1 \over c^2} \bigg\{
       2 {\Delta \over a^2} \Upsilon
       + 3 \left( \ddot U + 3 H \dot U
       + 2 {\ddot a \over a} U \right)
       - {2 \over a^2} U \Delta U
   \nonumber \\
   & & \qquad
       + {1 \over a^2} ( a P^i_{\;\;,i} )^{\displaystyle{\cdot}}
       + 8 \pi G \left[ {1 \over 2} ( \varrho \Pi
       - \varrho_b \Pi_b )
       + \varrho v^2
       + {3 \over 2} ( p - p_b ) \right] \bigg\} = 0,
   \label{PN-1-U} \\
   & & ( \dot U + H U )_{,i}
       + {1 \over 4 a} ( P^k_{\;\;,ki} - \Delta P_i )
       = 4 \pi G \varrho a v_i.
   \label{PN-2-U}
\eea

The energy and momentum conservation equations are
\bea
   & & - {1 \over c} T^b_{0;b}
       = \dot \varrho + 3 H \varrho
       + {1 \over a} ( \varrho v^i )_{,i}
       + {1 \over c^2} \bigg\{
       [ \varrho (\Pi + v^2 ) ]^{\displaystyle{\cdot}}
       + 3 H (\varrho \Pi + p )
   \nonumber \\
   & & \qquad
       + {1 \over a} \left[ \varrho v^i
       ( \Pi + v^2 + \Phi + 2 \Psi )
       + \varrho P^i + p v^i + Q^i + \Pi^i_j v^j \right]_{,i}
       + \varrho \left[ - 3 \dot \Psi
       + {1 \over a} ( \Phi - 3 \Psi )_{,i} v^i
       + 4 H v^2 \right]
       \bigg\},
   \label{E-conserv-PN} \\
   & & {1 \over a} T^b_{i;b}
       = {1 \over a^4} ( a^4 \varrho v_i )^{\displaystyle{\cdot}}
       + {1 \over a} \left( \varrho v^j v_i + p \delta^j_i
       + \Pi^j_i \right)_{,j}
       + {1 \over a} \varrho \Phi_{,i}
       + {1 \over c^2} \bigg\{
       {1 \over a^4} \left\{ a^4 \left[
       \varrho v_i ( \Pi + v^2 - \Phi ) + p v_i
       + Q_i + \Pi_{ij} v^j \right] \right\}^{\displaystyle{\cdot}}
   \nonumber \\
   & & \qquad
       + {1 \over a} \left[ \varrho v^j v_i
       ( \Pi + v^2 + 2 \Psi ) + \varrho P^j v_i
       + p v^j v_i + Q^j v_i + Q_i v^j
       + 2 \Psi \Pi^j_i \right]_{,j}
   \nonumber \\
   & & \qquad
       + ( \dot \Phi - 3 \dot \Psi ) \varrho v_i
       + {1 \over a} ( \Phi - 3 \Psi )_{,j}
       ( \varrho v^j v_i + \Pi^j_i )
       + {1 \over a} \Phi_{,i}
       \left[ \varrho ( \Pi + v^2 ) + p - 2 \Phi \right]
       + {1 \over a} \Psi_{,i} \varrho v^2
       + \varrho v^j P_{j,i}
       \bigg\}.
\eea
To the background order, from Eq.\ (\ref{E-conserv-PN}), we have
\bea
   & & \left[ \varrho_b
       \left( 1 + {\Pi_b \over c^2} \right) \right]^{\displaystyle{\cdot}}
       + 3 H \left[ \varrho_b
       \left( 1 + {\Pi_b \over c^2} \right)
       + {p_b \over c^2} \right] = 0.
   \label{BG-conservation}
\eea
Subtracting the background equation, and using $U$ and $\Upsilon$ we have
\bea
   & & ( \varrho - \varrho_b )^{\displaystyle{\cdot}}
       + 3 H ( \varrho - \varrho_b )
       + {1 \over a} ( \varrho v^i )_{,i}
       + {1 \over c^2} \bigg\{
       ( \varrho \Pi - \varrho_b \Pi_b + \varrho v^2 )^{\displaystyle{\cdot}}
       + 3 H (\varrho \Pi - \varrho_b \Pi_b + p - p_b )
   \nonumber \\
   & & \qquad
       + {1 \over a} \left[ \varrho v^i
       ( \Pi + v^2 - 3 U )
       + \varrho P^i + p v^i + Q^i + \Pi^i_j v^j \right]_{,i}
       + \varrho \left( 3 \dot U
       + {2 \over a} U_{,i} v^i
       + 4 H v^2 \right)
       \bigg\} = 0,
   \label{E-conserv-fluid} \\
   & & {1 \over a^4} ( a^4 \varrho v_i )^{\displaystyle{\cdot}}
       + {1 \over a} \left( \varrho v^j v_i + p \delta^j_i
       + \Pi^j_i \right)_{,j}
       - {1 \over a} \varrho U_{,i}
       + {1 \over c^2} \bigg\{
       {1 \over a^4} \left\{ a^4 \left[
       \varrho v_i ( \Pi + v^2 + U ) + p v_i
       + Q_i + \Pi_{ij} v^j \right] \right\}^{\displaystyle{\cdot}}
   \nonumber \\
   & & \qquad
       + {1 \over a} \left[ \varrho v^j v_i
       ( \Pi + v^2 - 2 U ) + \varrho P^j v_i
       + p v^j v_i + Q^j v_i + Q_i v^j
       - 2 U \Pi^j_i \right]_{,j}
   \nonumber \\
   & & \qquad
       + 2 \dot U \varrho v_i
       + {2 \over a} U_{,j}
       ( \varrho v^j v_i + \Pi^j_i )
       - {2 \over a} \varrho \Upsilon_{,i}
       - {1 \over a} U_{,i}
       \left[ \varrho ( \Pi + 2 v^2 ) + p \right]
       + \varrho v^j P_{j,i}
       \bigg\} = 0.
   \label{Mom-conserv-fluid}
\eea
For a general fluid to 1PN order, the energy and momentum conservation equations are in Eqs.\ (\ref{E-conserv-fluid}) and (\ref{Mom-conserv-fluid}), and Einstein's equation provides Eqs.\ (\ref{PN-1-U}) and (\ref{PN-2-U}). These provide a complete set of equations valid to 1PN order without imposing the temporal gauge condition; see below Eq.\ (\ref{PN-gauges}) for gauge conditions. The background evolution is described by Eqs.\ (\ref{BG-eqs}) and (\ref{BG-conservation}).

%%%%%%%%%%%%%%%%%%%%%%%%%%%%%%%%%%%%%%%%%%%%%%%%%%%%%%%%%%%%%%%%%
\subsection{1PN Schr\"odinger formulation}
                                      \label{sec:1PN-Schrodinger}

To 1PN order, Eq.\ (\ref{relativistic-Schrodinger-eq}) gives
\bea
   & & \left( {\Delta \over a^2}
       - 8 \pi \ell_s | \psi |^2 \right) \psi
       + {2 i m \over \hbar} \left( \dot \psi
       + {3 \over 2} H \psi \right)
       - {2 m^2 \over \hbar^2} \Phi \psi
       + {1 \over c^2} \bigg[
       - \ddot \psi - 3 H \dot \psi
       + 2 \Psi {\Delta \over a^2} \psi
       + {1 \over a^2} \left( \Phi - \Psi \right)^{,i} \psi_{,i}
   \nonumber \\
   & & \qquad
       + {2 i m \over \hbar}
       \left( - 2 \Phi \dot \psi
       + {1 \over a} P^i \psi_{,i} \right)
       + {i m \over \hbar} \left( {1 \over a} P^i_{\;\;,i}
       - \dot \Phi - 3 \dot \Psi - 6 H \Phi \right) \psi
       + {4 m^2 \over \hbar^2} \Phi^2 \psi
       \bigg]
       = 0.
   \label{Schrodinger-1PN}
\eea
Using the PN notation in Eq.\ (\ref{Chandrasekhar-notation}), we have
\bea
   & & \left( {\Delta \over a^2}
       - 8 \pi \ell_s | \psi |^2 \right) \psi
       + {2 i m \over \hbar} \left( \dot \psi
       + {3 \over 2} H \psi \right)
       + {2 m^2 \over \hbar^2} U \psi
       + {1 \over c^2} \bigg[
       - \ddot \psi - 3 H \dot \psi
       - 2 U {\Delta \over a^2} \psi
   \nonumber \\
   & & \qquad
       + {2 i m \over \hbar} \left( 2 U \dot \psi
       + {1 \over a} P^i \psi_{,i} \right)
       + {i m \over \hbar}
       \left( {1 \over a} P^i_{\;\;,i}
       + 4 \dot U + 6 H U \right) \psi
       + {2 m^2 \over \hbar^2}
       \left( 2 \Upsilon + U^2 \right) \psi
       \bigg]
       = 0.
   \label{Schrodinger-1PN-U}
\eea

Using the four-vector in Eq.\ (\ref{four-vector-1PN}), Eq.\ (\ref{fluid-K}) gives the energy-frame condition, $q_i \equiv 0$,
\bea
   & & m |\psi|^2 a v_i
       = - i \hbar {1 \over 2}
       \left( \psi_{,i} \psi^* - \psi^*_{,i} \psi \right)
       + {1 \over c^2} \bigg\{
       m |\psi|^2 \left( v^2 + \Phi \right) a v_i
       - i \hbar \left(
       \dot \psi \psi^* - \dot \psi^* \psi \right) a v_i
   \nonumber \\
   & & \qquad
       - {\hbar^2 \over 2 m}
       \left[ \psi_{,i} \dot \psi^* + \psi^*_{,i} \dot \psi
       + {1 \over a} \left( \psi_{,i} \psi^*_{,j}
       + \psi^*_{,i} \psi_{,j} \right) v^j
       \right]
       \bigg\},
\eea
and the fluid quantities
\bea
   & & \mu = m |\psi|^2 c^2
       + {\hbar^2 \over 2 m} \left\{
       {1 \over a^2} \psi^{,i} \psi^*_{,i}
       + {1 \over 2 |\psi|^2 a^2}
       \left( \psi^{,i} \psi^* - \psi^{*,i} \psi \right)
       \left( \psi_{,i} \psi^* - \psi^*_{,i} \psi \right)
       - {1 \over 2 a^2} \left[ ( \Delta \psi ) \psi^*
       + ( \Delta \psi^* ) \psi \right]
       + 12 \pi \ell_s | \psi |^4 \right\},
   \nonumber \\
   & & p = {\hbar^2 \over 2 m} \left\{
       - {1 \over 3 a^2} \psi^{,i} \psi^*_{,i}
       + {1 \over 6 |\psi|^2 a^2}
       \left( \psi^{,i} \psi^* - \psi^{*,i} \psi \right)
       \left( \psi_{,i} \psi^* - \psi^*_{,i} \psi \right)
       - {1 \over 2 a^2} \left[ ( \Delta \psi ) \psi^*
       + ( \Delta \psi^* ) \psi \right]
       + 4 \pi \ell_s | \psi |^4 \right\},
   \nonumber \\
   & & \qquad
       + {\hbar^2 \over 12 m c^2} \bigg\{
       6 |\dot \psi|^2
       - {4 \over a^2} \psi^{,i} \psi^*_{,i} \Psi
       + {2 \over a} ( \dot \psi \psi^*_{,i}
       + \dot \psi^* \psi_{,i} ) v^i
   \nonumber \\
   & & \qquad
       + 3 \left[ \ddot \psi + 3 H \dot \psi
       - 2 \Psi {\Delta \over a^2} \psi
       - {1 \over a^2} ( \Phi - \Psi )^{,i} \psi_{,i}
       \right] \psi^*
       + 3 \left[ \ddot \psi^* + 3 H \dot \psi^*
       - 2 \Psi {\Delta \over a^2} \psi^*
       - {1 \over a^2} ( \Phi - \Psi )^{,i} \psi^*_{,i}
       \right] \psi \bigg\}
   \nonumber \\
   & & \qquad
       + {\hbar^2 \over 6 m c^2 |\psi|^2} {1 \over a^2}
       \left\{
       a ( \dot \psi \psi^* - \dot \psi^* \psi )
       ( \psi_{,i} \psi^* - \psi^*_{,i} \psi ) v^i
       + {1 \over 2} \left[
       ( \psi_{,i} \psi^* - \psi^*_{,i} \psi ) v^i \right]^2
       + \Psi
       ( \psi^{,i} \psi^* - \psi^{*,i} \psi )
       ( \psi_{,i} \psi^* - \psi^*_{,i} \psi ) \right\}
   \nonumber \\
   & & \qquad
       - {i \hbar^3 \over 12 m^2 c^2 |\psi|^2} {1 \over a^2}
       ( \psi^{,i} \psi^* - \psi^{*,i} \psi )
       \left[
       ( \psi_{,i} \dot \psi^* + \psi^*_{,i} \dot \psi )
       + {1 \over a}
       ( \psi_{,i} \psi^*_{,j} + \psi^*_{,i} \psi_{,j} ) v^j
       \right], \quad
       Q_i \equiv 0,
   \nonumber
\eea
\bea
   & & \Pi_{ij}
       = {\hbar^2 \over m a^2} \bigg\{
       \psi_{,(i} \psi^*_{,j)}
       + {1 \over 4 |\psi|^2}
       ( \psi_{,i} \psi^* - \psi^*_{,i} \psi )
       ( \psi_{,j} \psi^* - \psi^*_{,j} \psi )
   \nonumber \\
   & & \qquad
       - {1 \over 3} \delta_{ij} \left[
       \psi^{,k} \psi^*_{,k}
       + {1 \over 4 |\psi|^2}
       \left( \psi^{,k} \psi^* - \psi^{*,k} \psi \right)
       \left( \psi_{,k} \psi^* - \psi^*_{,k} \psi \right)
       \right]
       \bigg\}
   \nonumber \\
   & & \qquad
       + {\hbar^2 \over m c^2} \bigg\{
       {1 \over a} ( \dot \psi + {1 \over a} \psi_{,k} v^k )
       v_{(i} \psi^*_{,j)}
       + {1 \over a} ( \dot \psi^* + {1 \over a} \psi^*_{,k} v^k )
       v_{(i} \psi_{,j)}
       - {1 \over 3 a^2} \psi^{,k} \psi^*_{,k} v_i v_j
   \nonumber \\
   & & \qquad
       - {1 \over 3} \delta_{ij}
       \left[ {1 \over a} ( \dot \psi \psi^*_{,k}
       + \dot \psi^* \psi_{,k} ) v^k
       + {1 \over a^2} | \psi_{,k} v^k |^2 \right] \bigg\}
   \nonumber \\
   & & \qquad
       + {\hbar^2 \over m c^2 |\psi|^2} \bigg\{
       {1 \over 2 a} ( \psi_{,(i} \psi^*
       - \psi^*_{,(i} \psi ) v_{j)}
       \left[ \dot \psi \psi^* - \dot \psi^* \psi
       + {1 \over a} ( \psi_{,k} \psi^*
       - \psi^*_{,k} \psi ) v^k \right]
   \nonumber \\
   & & \qquad
       - {1 \over 12 a^2}
       \left( \psi^{,k} \psi^* - \psi^{*,k} \psi \right)
       \left( \psi_{,k} \psi^* - \psi^*_{,k} \psi \right)
       v_i v_j
   \nonumber \\
   & & \qquad
       - {1 \over 12} \delta_{ij}
       \left[ {2 \over a} ( \dot \psi \psi^* - \dot \psi^* \psi )
       ( \psi_{,k} \psi^* - \psi^*_{,k} \psi ) v^k
       + {1 \over a^2} \left[
       ( \psi_{,k} \psi^* - \psi^*_{,k} \psi ) v^k \right]^2
       \right] \bigg\}.
   \label{fluid-1PN-K}
\eea

Using the axion fluid quantities, Einstein's equation in (\ref{PN-1-U}) and (\ref{PN-2-U}) give
\bea
   & & {\Delta \over a^2} U
       + 4 \pi G m ( |\psi|^2 - |\psi_b|^2 )
       + {1 \over c^2} \bigg\{
       2 {\Delta \over a^2} \Upsilon
       + 3 \left( \ddot U + 3 H \dot U
       + 2 {\ddot a \over a} U \right)
       - {2 \over a^2} U \Delta U
       + {1 \over a^2} ( a P^i_{\;\;,i} )^{\displaystyle{\cdot}}
   \nonumber \\
   & & \qquad
       + 8 \pi G {\hbar^2 \over m}
       \left[ - {1 \over 2 a^2} \left( \psi \Delta \psi^*
       + \psi^* \Delta \psi \right)
       + 6 \pi \ell_s ( |\psi|^4 - |\psi_b|^4 )
       \right] \bigg\} = 0,
   \label{PN-1-U-axion-Schrodinger} \\
   & & ( \dot U + H U )_{,i}
       + {1 \over 4 a} ( P^k_{\;\;,ki} - \Delta P_i )
       = 2 \pi G i \hbar \left( \psi \psi^*_{,i}
       - \psi^* \psi_{,i} \right).
   \label{PN-2-U-axion-Schrodinger}
\eea
Equations (\ref{Schrodinger-1PN-U}), (\ref{PN-1-U-axion-Schrodinger}) and (\ref{PN-2-U-axion-Schrodinger}) provide a complete set of axion fluid to 1PN order without imposing the temporal gauge condition, see below Eq.\ (\ref{PN-gauges}) for gauge conditions. The background evolution is described by Eqs.\ (\ref{BG-eqs}) and (\ref{BG-conservation}), with
\bea
   & & \varrho_b = m |\psi_b|^2, \quad
       \varrho_b \Pi_b = 3 p_b
       = {6 \pi \ell_s \hbar^2 \over m} |\psi_b|^4.
\eea

%%%%%%%%%%%%%%%%%%%%%%%%%%%%%%%%%%%%%%%%%%%%%%%%%%%%%%%%%%%%%%%%%
\subsection{1PN Madelung formulation}
                                      \label{sec:1PN-Madelung}

Under the Madelung transformation
the imaginary and real parts of Eq.\ (\ref{Schrodinger-1PN}), or directly from Eqs.\ (\ref{imaginary-Klein}) and (\ref{real-Klein}), give
\bea
   & & \dot \varrho + 3 H \varrho
       + {1 \over a^2} \left( \varrho u^{,i} \right)_{,i}
   \nonumber \\
   & & \qquad
       + {1 \over c^2} \left[
       - \left( \varrho \dot u \right)^{\displaystyle{\cdot}}
       - 3 H \varrho \dot u
       + 2 \left( \Phi + \Psi \right)
       {1 \over a^2} \left( \varrho u^{,i} \right)_{,i}
       + {1 \over a} \left( \varrho P^i \right)_{,i}
       - \varrho \left( \dot \Phi + 3 \dot \Psi \right)
       + {1 \over a^2} \varrho \left( \Phi - \Psi \right)^{,i} u_{,i}
       \right]
       = 0,
   \\
   & & \dot u + {1 \over 2 a^2} u^{,i} u_{,i}
       + \Phi - {\hbar^2 \over 2 m^2}
       \left( {1 \over a^2}
       {\Delta \sqrt{\varrho} \over \sqrt{\varrho}}
       - {8 \pi \ell_s \over m} \varrho \right)
       + {1 \over c^2} \bigg\{ - {1 \over 2} \dot u^2
       + \left( \Phi + \Psi \right)
       {1 \over a^2} u^{,i} u_{,i}
       + {1 \over a} P^i u_{,i}
   \nonumber \\
   & & \qquad
       + {\hbar^2 \over 2 m^2} \bigg[
       {\ddot {\sqrt{\varrho}} \over \sqrt{\varrho}}
       + 3 H {\dot {\sqrt{\varrho}} \over \sqrt{\varrho}}
       - 2 \left( \Phi + \Psi \right)
       {1 \over a^2} {\Delta \sqrt{\varrho} \over \sqrt{\varrho}}
       + \Phi {16 \pi \ell_s \over m} \varrho
       - {1 \over a^2} \left( \Phi - \Psi \right)^{,i}
       {\sqrt{\varrho}_{,i} \over \sqrt{\varrho}}
       \bigg] \bigg\}
       = 0.
\eea
By identifying ${\bf u} \equiv {1 \over a} \nabla u$, we have
\bea
   & & \dot \varrho + 3 H \varrho
       + {1 \over a} \nabla \cdot \left( \varrho {\bf u} \right)
       + {1 \over c^2} \bigg[
       - \left( \varrho \dot u \right)^{\displaystyle{\cdot}}
       - 3 H \varrho \dot u
       - \varrho \left( \dot \Phi + 3 \dot \Psi \right)
   \nonumber \\
   & & \qquad
       + 2 \left( \Phi + \Psi \right) {1 \over a}
       \nabla \cdot \left( \varrho {\bf u} \right)
       + {1 \over a} \varrho {\bf u} \cdot \nabla
       \left( \Phi - \Psi \right)
       + {1 \over a} \nabla \cdot \left( \varrho {\bf P} \right)
       \bigg]
       = 0,
   \label{Continuity-1PN-KM} \\
   & & \dot {\bf u} + H {\bf u}
       + {1 \over a} {\bf u} \cdot \nabla {\bf u}
       + {1 \over a} \nabla \Phi
       - {\hbar^2 \over 2 m^2}
       {1 \over a} \nabla \left( {1 \over a^2}
       {\Delta \sqrt{\varrho} \over \sqrt{\varrho}}
       - {8 \pi \ell_s \over m} \varrho \right)
       + {1 \over c^2} {1 \over a} \nabla \bigg\{
       - {1 \over 2} \dot u^2
       + \left( \Phi + \Psi \right) {\bf u}^2
       + {\bf u} \cdot {\bf P}
   \nonumber \\
   & & \qquad
       + {\hbar^2 \over 2 m^2} \bigg[
       {\ddot {\sqrt{\varrho}} \over \sqrt{\varrho}}
       + 3 H {\dot {\sqrt{\varrho}} \over \sqrt{\varrho}}
       - 2 \left( \Phi + \Psi \right)
       {1 \over a^2}
       {\Delta \sqrt{\varrho} \over \sqrt{\varrho}}
       + \Phi {16 \pi \ell_s \over m} \varrho
       - {1 \over a^2} {1 \over \sqrt{\varrho}} (\nabla \sqrt{\varrho}) \cdot
       \nabla \left( \Phi - \Psi \right)
       \bigg] \bigg\}
       = 0,
   \label{Euler-1PN-KM}
\eea
and, for remaining $\dot u$, we can use
\bea
   & & \dot u = - \Phi
       - {1 \over 2} {\bf u}^2
       + {\hbar^2 \over 2 m^2}
       \left( {1 \over a^2}
       {\Delta \sqrt{\varrho} \over \sqrt{\varrho}}
       - {8 \pi \ell_s \over m} \varrho \right).
   \label{Euler-0PN-KM}
\eea
Using Chandrasekhar's 1PN notation in Eq.\ (\ref{Chandrasekhar-notation}) we have
\bea
   & & \dot \varrho + 3 H \varrho
       + {1 \over a} \nabla \cdot \left( \varrho {\bf u} \right)
       + {1 \over c^2} \left[
       - \left( \varrho \dot u \right)^{\displaystyle{\cdot}}
       - 3 H \varrho \dot u
       - 4 U {1 \over a} \nabla \cdot \left( \varrho {\bf u} \right)
       + 4 \varrho \dot U
       + {1 \over a} \nabla \cdot \left( \varrho {\bf P} \right)
       \right]
       = 0,
   \label{E-conserv-axion} \\
   & & \dot {\bf u} + H {\bf u}
       + {1 \over a} {\bf u} \cdot \nabla {\bf u}
       - {1 \over a} \nabla U
       - {\hbar^2 \over 2 m^2} {1 \over a} \nabla
       \left( {1 \over a^2} {\Delta \sqrt{\varrho} \over \sqrt{\varrho}}
       - {8 \pi \ell_s \over m} \varrho \right)
   \nonumber \\
   & & \qquad
       + {1 \over c^2} {1 \over a}
       \nabla \left[ - {1 \over 2} \dot u^2
       - 2 \Upsilon + U^2
       - 2 U {\bf u}^2
       + {\bf P} \cdot {\bf u}
       + {\hbar^2 \over 2 m^2} \left(
       {\ddot {\sqrt{\varrho}} \over \sqrt{\varrho}}
       + 3 H {\dot {\sqrt{\varrho}} \over \sqrt{\varrho}}
       + 4 U {1 \over a^2}
       {\Delta \sqrt{\varrho} \over \sqrt{\varrho}}
       - U {16 \pi \ell_s \over m} \varrho
       \right) \right]
       = 0,
   \label{Mom-conserv-axion} \\
   & & \dot u = U
       - {1 \over 2} {\bf u}^2
       + {\hbar^2 \over 2 m^2}
       \left( {1 \over a^2}
       {\Delta \sqrt{\varrho} \over \sqrt{\varrho}}
       - {8 \pi \ell_s \over m} \varrho \right).
   \label{dot-u-0PN}
\eea

To 1PN order, using the four-vector in Eq.\ (\ref{four-vector-1PN}), Eq.\ (\ref{fluid-KM}) gives the fluid quantities. The energy-frame condition, $q_a \equiv 0$, in Eq.\ (\ref{energy-frame-KM}) gives
\bea
   & & u_{,i} = a v_i \left[ 1 - {1 \over c^2} \left(
       \dot u + \Phi \right) \right]
       + {\hbar^2 \over m^2 c^2}
       \left( {\dot{\sqrt{\varrho}} \over \sqrt{\varrho}}
       + {1 \over a}
       {\sqrt{\varrho}_{,k} \over \sqrt{\varrho}} v^k \right)
       {\sqrt{\varrho}_{,i} \over \sqrt{\varrho}}.
   \label{u-v-relation}
\eea
Thus, to 0PN order $u_{,i} = a v_i$. Using the notation in Eq.\ (\ref{fluid-PN}), the fluid quantities follow from Eq.\ (\ref{fluid-KM}). We can identify $\varrho$ as the mass density, and have
\bea
   & & \varrho \Pi = {\hbar^2 \over 2 m^2} \left( {1 \over a^2}
       \sqrt{\varrho}^{,i} \sqrt{\varrho}_{,i}
       - {1 \over a^2} \sqrt{\varrho} \Delta \sqrt{\varrho}
       + {12 \pi \ell_s \over m} \varrho^2 \right),
   \nonumber \\
   & & p = {\hbar^2 \over 6 m^2} \bigg\{
       - {1 \over a^2}
       \sqrt{\varrho}^{,i} \sqrt{\varrho}_{,i}
       - {3 \over a^2} \sqrt{\varrho} \Delta \sqrt{\varrho}
       + {12 \pi \ell_s \over m} \varrho^2
       + {1 \over c^2} \bigg[
       3 \sqrt{\varrho} \left( \ddot {\sqrt{\varrho}}
       + 3 H \dot {\sqrt{\varrho}} \right)
       + 3 ( \dot {\sqrt{\varrho}} )^2
   \nonumber \\
   & & \qquad
       + 4 \dot {\sqrt{\varrho}}
       {1 \over a} \sqrt{\varrho}_{,i} v^i
       + {2 \over a^2} ( \sqrt{\varrho}_{,i} v^i )^2
       - {2 \over a^2} \Psi \left(
       \sqrt{\varrho}^{,i} \sqrt{\varrho}_{,i}
       + 3 \sqrt{\varrho} \Delta \sqrt{\varrho} \right)
       - {3 \over a^2} ( \Phi - \Psi )^{,i}
       \sqrt{\varrho} \sqrt{\varrho}_{,i} \bigg] \bigg\}, \quad
       Q_i \equiv 0,
   \nonumber \\
   & & \Pi_{ij} = {\hbar^2 \over m^2 a^2} \bigg\{
       \sqrt{\varrho}_{,i} \sqrt{\varrho}_{,j}
       - {1 \over 3} \sqrt{\varrho}^{,k} \sqrt{\varrho}_{,k}
       \delta_{ij}
       + {1 \over c^2} \bigg[
       a \left( \dot {\sqrt{\varrho}}
       + {1 \over a} \sqrt{\varrho}_{,k} v^k \right)
       \left( \sqrt{\varrho}_{,i} v_j + \sqrt{\varrho}_{,j} v_i
       - {2 \over 3} \delta_{ij} \sqrt{\varrho}_{,k} v^k \right)
   \nonumber \\
   & & \qquad
       + {1 \over 3} \delta_{ij} ( \sqrt{\varrho}_{,k} v^k )^2
       - {1 \over 3} \sqrt{\varrho}^{,k} \sqrt{\varrho}_{,k}
       v_i v_j \bigg] \bigg\}.
   \label{fluid-axion-1PN}
\eea
This also follows from Eq.\ (\ref{fluid-1PN-K}). Notice that $p$ and $\Pi_{ij}$ are derived up to 2PN order as we need that order to derive the axion conservation equations in (\ref{E-conserv-axion}) and (\ref{Mom-conserv-axion}) from the fluid conservation equations in (\ref{E-conserv-fluid}) and (\ref{Mom-conserv-fluid}); we also need to use Eqs.\ (\ref{dot-u-0PN}) and (\ref{u-v-relation}).

Using the axion fluid quantities, Einstein's equations in (\ref{PN-1-U}) and (\ref{PN-2-U}) give
\bea
   & & {\Delta \over a^2} U
       + 4 \pi G ( \varrho - \varrho_b )
       + {1 \over c^2} \bigg\{
       2 {\Delta \over a^2} \Upsilon
       + 3 \left( \ddot U + 3 H \dot U
       + 2 {\ddot a \over a} U \right)
       - {2 \over a^2} U \Delta U
       + {1 \over a^2} ( a P^i_{\;\;,i} )^{\displaystyle{\cdot}}
   \nonumber \\
   & & \qquad
       + 8 \pi G \varrho u_i^2
       + 8 \pi G {\hbar^2 \over m^2}
       \left[ - {1 \over a^2} \sqrt{\varrho} \Delta \sqrt{\varrho}
       + {6 \pi \ell_s \over m} (\varrho^2 - \varrho_b^2)
       \right] \bigg\} = 0,
   \label{PN-1-U-axion} \\
   & & ( \dot U + H U )_{,i}
       + {1 \over 4 a} ( P^k_{\;\;,ki} - \Delta P_i )
       = 4 \pi G \varrho a u_i.
   \label{PN-2-U-axion}
\eea
These also follow from Eqs.\ (\ref{PN-1-U-axion-Schrodinger}) and (\ref{PN-2-U-axion-Schrodinger}) using the Madelung transformation.

Equations (\ref{E-conserv-axion}), (\ref{Mom-conserv-axion}), (\ref{PN-1-U-axion}) and (\ref{PN-2-U-axion}) provide a complete set of axion in Schr\"odinger formulation to 1PN order without fixing the temporal gauge condition; see below Eq.\ (\ref{PN-gauges}) for gauge conditions. The background evolution is described by Eqs.\ (\ref{BG-eqs}) and (\ref{BG-conservation}), with
\bea
   & & \varrho_b \Pi_b = 3 p_b
       = {6 \pi \ell_s \hbar^2 \over m^3} \varrho_b^2.
\eea

\end{widetext}
%%%%%%%%%%%%%%%%%%%%%%%%%%%%%%%%%%%%%%%%%%%%%%%%%%%%%%%%%%%%%%%%%
\end{document}